\documentclass[preprint2]{pp7}

\usepackage{xcolor}
\renewcommand\fbox{\fcolorbox{red}{white}}

\input{pp7.h}

\begin{document}

\title{\textbf{\LARGE The Solar Neighborhood in the Age of Gaia
}}

\author {\textbf{\large Catherine Zucker\footnote{NASA Hubble Fellow}}}
\affil{\small\it Space Telescope Science Institute, 3700 San Martin Dr, Baltimore, MD 21218, USA}
\author {\textbf{\large João Alves}}
\affil{\small\it University of Vienna, Department of Astrophysics, Türkenschanzstraße 17, 1180 Vienna, Austria}
\author {\textbf{\large Alyssa Goodman}}
\affil{\small\it Center for Astrophysics $\vert$ Harvard \& Smithsonian, 60 Garden St., Cambridge, MA, USA 02138}
\author {\textbf{\large Stefan Meingast}}
\affil{\small\it University of Vienna, Department of Astrophysics, Türkenschanzstraße 17, 1180 Vienna, Austria}
\author {\textbf{\large Phillip Galli}}
\affil{\small\it Núcleo de Astrofísica Teórica, Universidade Cidade de São Paulo, R. Galvão Bueno 868, Liberdade, 01506-000, São Paulo, SP, Brazil}

\begin{abstract}
\baselineskip = 11pt
\leftskip = 1.5cm 
\rightskip = 1.5cm
\parindent=1pc
{\small 
Most of what we know about the formation of stars, and essentially everything we know about the formation of planets, comes from observations of our solar neighborhood within 2 kpc of the Sun. Before 2018, accurate distance measurements needed to turn the 2D Sky into a faithful 3D physical picture of the distribution of stars, and the interstellar matter that forms them, were few and far between. Here, we offer a holistic review of how, since 2018, data from the \textit{Gaia} mission are revealing previously unseen and often unexpected 3D distributions of gas, dust, and young stars in the solar neighborhood. We summarize how new extinction-based techniques yield 3D dust maps and how the density structure mapped out offers key context for measuring young stars’ 3D positions from \textit{Gaia} and VLBI. We discuss how a subset of young stars in \textit{Gaia} with measured radial velocities and proper motions is being used to recover 3D cloud motion and characterize the internal dynamics of individual star-forming regions. We review relationships between newly-identified clusters and streams of young stars and the molecular interstellar medium from which they evolve. The combination of these measures of gas and stars' 3D distribution and 3D motions provides unprecedented data for comparison with simulations and reframes our understanding of local star formation in a larger Galactic context. This new 3D view of our solar neighborhood in the age of \textit{Gaia} shows that star-forming regions once thought to be isolated are often connected on kiloparsec scales, causing us to reconsider models for the arrangement of gas and young stars in galaxies. 
}

\bigskip
\bigskip
\end{abstract}


\section{\textbf{INTRODUCTION}}

The local Milky Way offers the highest-resolution laboratory for studying the formation and evolution of planets, stars, and galaxies. 
Driven in large part by a combination of stellar birth and death, only within the local Milky Way is it possible to observe the entire lifecycle of star-forming interstellar gas, from kiloparsec scales down to au scales. 

Fortunately, even within just the nearest 2 kpc, we can probe a wide range of physical processes that influence the arrangement and kinematics of interstellar gas, dust, and young stars. On the largest scales, constraints on the inflow and outflow of gas at the disk-halo interface --- key drivers of how galaxies sustain and quench star formation over cosmic time --- are obtained from absorption line studies of diffuse, high-latitude gas in the solar neighborhood \citep{Bish2019, Marasco2013}. In the same vein, current understanding of how the Milky Way's star formation history may be shaped by galaxy interactions and mergers \citep[e.g., with the Gaia-Sausage-Enceladus satellite or the Sagittarius dwarf galaxy;][]{Xiang2022, Ruiz-Lara2020} are primarily based on stellar phase-space studies of the local Milky Way. 

Within the disk, the solar neighborhood intersects a significant swath of arm and interarm structure \citep{Reid2019} and is interspersed with several feedback-driven loops, shells, and superbubbles \citep{Frisch2018}. It harbors dozens of molecular cloud complexes, both emerging and dispersing, hosting both low- and high-mass star formation \citep{Reipurth2008a,Reipurth2008b}. Within these cloud complexes, the solar neighborhood enables a census of pre-stellar cores\index{dense cores} and protostars\index{protostars}, allowing for characterization of the shape of the core\index{core mass function} and initial stellar mass function \citep[e.g.,][]{Alves2007-id,Andre2010,Bastian2010}. Moreover, as young clusters and associations emerge from their natal clouds, the solar neighborhood provides a backdrop for understanding how their dynamical evolution relates to the physical conditions of their birth. For all these reasons, the solar neighborhood allows us to understand the conditions of star formation across space and time, from the scale of spiral arms down to the scale of circumstellar disks, and from the youngest embedded protostars to older stellar associations dissolving into the galactic field. 

However, to fully understand the processes that shape star formation within the disk (e.g., gravity, turbulence, magnetic fields, feedback, and galactic dynamics), we need to build an interconnected physical picture of young stars and the gas and dust from which they form. Such a picture needs three dimensions of space (e.g. longitude $l$, latitude $b$, distance $d$) and three dimensions of velocity (e.g. the transverse velocities, $v_l$, $v_b$, and the radial velocity $v_R$) --- easy in simulation, but much harder in observation. Because plane-of-the-sky observations of the solar neighborhood have provided much of our understanding of star formation over the past century, most of what we think we know is based on 2D ($l$, $b$) or false ``3D” ($l$-$b$-$v_R$) observations, where the third dimension is the radial velocity obtained from the Doppler effect, not distance. Widespread, homogeneous constraints on distances, as well as transverse velocities, have largely been absent. Note that transverse velocities alone (e.g. $v_l$, $v_b$), now easily accessible in \textit{Gaia}, can be misleading. For example, star clusters or clouds at identical distances and with identical 3D Heliocentric Cartesian Space velocities (e.g. $U$, $V$, and $W$) can have varying $v_l$, $v_b$, $v_R$, due to simple geometric (projection) effects \citep[see e.g.][]{Meingast2021}.

In the last decades, star-based extinction maps\index{dust!extinction}, followed by \textit{Herschel} and \textit{Planck} maps, have given rise to exquisite 2D plane-of-the-sky maps of the dust column density toward nearby star-forming regions \citep{Lada1994,Lombardi2001, Lombardi2014,Andre2010,Arzoumanian2011, Dore2014, Kirk2014}. Yet distances to these clouds were not traditionally known to better than 20-30\% accuracy, hampering not only the determination of their basic physical properties (e.g., mass, size), but also our ability to characterize both their internal structure (e.g., thickness, gradients) and global distribution (e.g. relationship to spiral structure). Spectral-line maps of CO \citep{Dame2001} and HI \citep{HI4PI} in $l$-$b$-$v_R$ space have provided a critical third dimension, useful not only for disentangling the projected structure seen in 2D dust maps but also providing much needed kinematic constraints on phenomena like outflows and stellar winds \citep{Arce2010, Nishimura2015}. However, correspondence between intensity features (in $l$-$b$-$v_R$ space) and real density features (in $l$-$b$-$d$ space) has long been fraught with peril \citep{Beaumont2013}, hampering the ability of spectral-line maps alone to reveal the physical conditions of star formation in molecular clouds.

Likewise, infrared surveys like WISE \citep{Wright2010}, 2MASS \citep{Skrutskie2006}\index{2MASS}, and \textit{Spitzer} \citep{Gutermuth2008}\index{Spitzer Space Telescope} have enabled robust identification of the 2D plane-of-the-sky distribution of young stars — from deeply embedded systems to revealed OB associations — based on their infrared colors. However, radial velocities have typically only been available for a small subset of young stars. They were primarily obtained inhomogeneously via spectroscopic follow-up on a case-by-case basis \citep{Prieto2008, Szentgyorgyi2011, Torres2006}. Hipparcos provided constraints on the distance and transverse velocities of 100,000 stars in 1997 \citep{Perryman1997}, but not significantly beyond a few hundred parsecs and without the sensitivity to resolve any but the brightest young stellar objects. Without systematic constraints on the distances and 3D motions of young stars, it has traditionally been difficult to tell whether a given stellar association is a real physical entity (as opposed to a chance projection), let alone its dynamical state — whether it is bound, unbound, expanding, or contracting \citep{Mamajek2015}. 

Thanks to the launch of the \textit{Gaia} space mission in 2013\index{Gaia satellite}, this situation has changed \citep{Prusti2016}. The primary goal of the \textit{Gaia} mission is to unlock the three-dimensional spatial and three-dimensional velocity distributions of stars in the Milky Way. To do so, its main data product is astrometric measurements of stellar positions, parallaxes, and proper motions, as well as high-resolution spectrosopic measurements for the brightest stars. Together, these measurements provide unparalleled constraints on stellar distances, transverse velocities, and radial velocities across the full sky. 

The first \textit{Gaia} data release in 2016 (\textit{Gaia} DR1) provided positions, parallaxes, and proper motions for 2.5 million stars in common with the Hipparcos and Tycho-2 catalogs \citep{Brown2016}. The second data release of \textit{Gaia} in 2018 (\textit{Gaia} DR2) had a significantly larger data volume, providing positions, parallaxes, and proper motions for 1.3 billion stars. With a typical systematic sky-averaged astrometric uncertainty of $<$ 0.1 mas, \textit{Gaia} DR2 charted $10,000\times$ as many stars, with a $100\times$ better astrometric precision than its predecessor Hipparcos \citep{Gaia2018, Riello2018, Lindegren2018}. The third early data release in late 2020 (\textit{Gaia} EDR3) continued this trend, leveraging its more extended time baseline to improve parallax precisions by 30\% and proper motion precisions by a factor of two compared to \textit{Gaia} DR2 \citep{Lindegren2021, Riello2021}. The third data release in mid-2022, \textit{Gaia} DR3 \citep{Collaboration2022}, has provided astrophysical parameters \citep[e.g., surface gravity, effective temperature, metallicity;][]{Creevey2022, Fouesneau2022} for roughly half a billion stars based on lower-resolution $BP-RP$ spectra \citep{DeAngeli2022, Montegriffo2022}, as well as radial velocity measurements \citep{Katz2022} for an increased 33 million stars based on higher-resolution RVS spectra. More detailed astrophysical parameters based on the RVS spectra, including individual chemical abundances for 5 million stars, have also been provided in \textit{Gaia} DR3 using the GSP-Spec module \citep{Recio_Blanco_2022, Gaia_Collaboration_2022b}. \textit{Gaia} DR3 also released the first astrometric solutions for non-single stars \citep{Halbwachs2022}. 

\textit{Gaia}\index{Gaia satellite} alone can recover 5D information (e.g. $l$, $b$, $d$, $v_l$, $v_b$) for more than a billion stars, and 6D information (e.g. $l$, $b$, $d$, $v_l$, $v_b$, $v_R$) for a subset of the brightest stars. When combining \textit{Gaia}’s astrometry\index{Astrometry} with complementary spectroscopic information on the radial velocities of stars --- both from \textit{Gaia} and ancillary surveys such as SDSS, GALAH, LAMOST, and RAVE \citep{APOGEE, GALAH, LAMOST, RAVE} --- \textit{Gaia} is fulfilling its mission of constructing a 6D physical picture of the Milky Way’s stellar content, including a sizeable sample of young stars \citep{Gagne2018a, Galli2020-Lupus, Grosschedl2021, Ortiz-Leon2018_Serpens}. However, \textit{Gaia} alone does not provide information on the 3D structure of the interstellar medium. Thankfully, the simultaneous rise of not only \textit{Gaia} but also deep wide-field photometric surveys \citep[e.g., Pan-STARRS1;][]{Chambers2016} and advanced statistical techniques has allowed the field of 3D dust mapping to flourish, rendering never-before-seen 3D spatial maps of the interstellar medium \citep{Leike2020,Lallement2022,Green2019}. 

In this review, we argue that combining the new 6D spatial and velocity information on young stars and the new 3D spatial information on interstellar matter obtained from 3D dust mapping provides the foundation for constructing a physical picture of star formation previously attainable only in simulations. Ultimately, \textit{Gaia} is enabling us to bridge the gap: between 2D plane-of-the-sky views and true 3D space, between radial velocities and full 3D space motion, between different tracers of the interstellar medium (young stars, gas, and dust), across different size scales (parsecs to kiloparsecs), and as a function of time and evolutionary state.

In \S\ref{section2}, we summarize what \textit{Gaia} has taught us about the 3D structure of the local interstellar medium from large to small scales. We review recent measurements of distances to star-forming regions within 2 kpc of the Sun, derived from 3D dust mapping and trigonometric parallax observations of young stellar objects with \textit{Gaia} and VLBI. We highlight how these techniques, and combinations of them, are being used to probe the general topological distribution of gas and dust on large scales as well as to reveal clouds' internal 3D structure.

In \S\ref{section3}, we summarize how \textit{Gaia} has improved our understanding of the 6D structure of star-forming regions, constraining not only the 3D positions of stars but critically also their 3D velocities. Focusing on the more embedded stellar populations, which most likely trace the natal gas motion, we will show how the state-of-the-art astrometry (from \textit{Gaia} and VLBI) can be combined with ancillary radial velocity surveys to derive the 3D space motions of young stars, trace their dynamical evolution in time, and reconstruct the local history of star formation. 

In \S\ref{section4}, we explore the topological, temporal, and kinematic relationships between young stars and the molecular interstellar medium from which they form, emphasizing the role of feedback and wind-blown bubbles on star formation in the solar neighborhood. We then review the impact \textit{Gaia} has had so far on our understanding of the 3D distribution of young stars near the Sun and the identification of new clusters, streams, and aggregates.

In \S\ref{section5}, we will summarize the evidence for the existing large-scale features and present discoveries based on the unprecedented accuracy of the \textit{Gaia} mission, showing that local star-forming regions are much less isolated, or randomly distributed, than previously thought. We will also review existing theories for the global architecture of gas and young stars in the solar neighborhood in relation to our current understanding of Galactic structure and dynamics. 

Finally, in \S\ref{conclusions}, we summarize the key takeaways of this chapter and briefly look toward the future. 

\section{\textbf{DISTANCES AND TOPOLOGIES OF LOCAL MOLECULAR CLOUDS}} \label{section2}

Accurate distance estimates to local star-forming regions in the post-\textit{Gaia} era broadly come in two flavors: statistical measures of the 3D dust distribution and parallax measurements of young stellar objects embedded in the clouds themselves\index{Giant molecular clouds}. 

As discussed in \S\ref{3ddust}, 3D dust mapping traces the overall 3D spatial structure of the interstellar medium by using inference pipelines that leverage stellar photometric surveys to model the line-of-sight distribution of dust that reddens star colors \citep{Chen2018, Green2019, Guo2021, Lallement2019}. 3D dust mapping extends back to the pre-\textit{Gaia} era \citep{Green2015, Lallement2003, Marshall2006}, but has recently received a major distance resolution boost because of the constraints on stellar distances (independent of their colors) that \textit{Gaia} provides. In the \textit{Gaia}-era, even-more-optimized versions of 3D dust mapping that sacrifice spatial for distance resolution can infer dust-based distances to specific molecular clouds with accuracies of $\approx$ 5\% out to at least 2 kpc \citep{Chen2020b, Yan2021a, Zucker2019, Zucker2020, Yan2019a, Yan2019b}. And 3D dust maps that trade off volume coverage of the Galaxy for increased distance resolution nearby --- mainly covering the nearest few hundred parsecs --- can achieve the highest distance resolution, on the order of 1 pc, enabling fundamental new measurements of cloud’s internal structure \citep{Leike2020, Leike2019}. 

As discussed in \S\ref{stellar_distances}, radio-frequency surveys like GOBELINS, BeSSeL, or VERA \citep[e.g.][]{Loinard2012,Reid2019} complement 3D dust mapping techniques and can measure distances using geometric parallax for deeply-embedded young stars (and/or associated masers), while \textit{Gaia} astrometry gives accurate distances to less-extinguished class II and III protostars \index{Class II Sources} \index{Class II Sources} \citep[e.g.][]{Grosschedl2018}. 

Finally, as discussed in \S\ref{agreement}, we consider the broad agreement between 3D dust mapping and YSO parallax-based distances in the solar neighborhood and show how they can be combined to provide precise distance constraints to sub-regions of the clouds over a large dynamic range in density, from dense star-forming clumps to diffuse molecular cloud envelopes. 

\subsection{\textbf{3D Dust Mapping of the Local ISM}}\label{3ddust}
\subsubsection{\textbf{3D Dust Mapping and the Rise of Gaia}} 

Using 3D dust mapping to construct a 3D spatial view of the interstellar medium relies on the principle that dust reddens starlight. The dust associated with a cloud will both extinguish and redden the light from stars in proportion to the column density of dust along the line of sight to each star. Using advanced statistical and computational techniques, 3D dust mapping models the observed distribution of stellar photometric colors on the 2D Sky as produced by a self-consistent 3D distribution of dust, stellar types, and stellar distances. Because young stellar objects are challenging to model and relatively scarce, 3D dust mapping models the photometry of the main sequence stars foreground and background to the dust cloud, which is publicly and cheaply available for billions of stars across the Milky Way. Since gas and dust are highly correlated, constraining the 3D spatial distribution of dust also enables constraints on the spatial distribution of interstellar gas, including nearby star-forming regions. 

Before \textit{Gaia} DR2, astrometric constraints on stars’ distances independent of their colors were scarce, making it challenging to construct accurate 3D models of the interstellar medium using photometry alone. While several such ground-breaking 3D dust maps were available in the pre-\textit{Gaia} era \citep{Green2015, Lallement2003, Lombardi2001, Marshall2006, Sale2014}, at best, the 3D spatial resolution of such maps was several times larger than the typical thicknesses of clouds. By placing astrometric constraints on the distance to over a billion stars, \textit{Gaia} has also produced extraordinary gains in resolving the 3D dust distribution essentially overnight, reducing distance uncertainties from 20\% in the pre-\textit{Gaia} era \citep{Green2015}, down to $1-5\%$ in the present day \citep{Leike2020}. And the fidelity of 3D dust maps will only continue to improve as more and more sensitive photometric surveys are combined with the astrometry from current and future \textit{Gaia} data releases. 

While the principle of 3D dust mapping remains the same across the literature, the distance resolution and coverage of different approaches vary significantly. Because 3D dust maps are computationally expensive to produce, maps that target a larger fraction of the Galaxy (both in distance and on the plane of the sky) tend to have lower distance resolution than maps with smaller coverage. We discuss three different 3D-dust-based data products in order of increasing distance resolution. 

First, in \S\ref{grossdust}, we start with maps that chart the gross distribution ($\approx$10\% distance uncertainty) of the interstellar medium over a range of densities out to at least 2 kpc. Next, in \S\ref{targeteddust} we review complementary techniques that only constrain distance to star-forming molecular gas over the same extent, but with 2$\times$ higher distance uncertainty ($\approx$ 5\%). Finally, in \S\ref{duststructure}, we discuss maps that achieve the highest resolution --- capable of revealing the internal structure of clouds at $\approx$1\% distance uncertainty --- but are largely restricted to the nearest few hundred parsecs and/or small volume cutouts around select molecular clouds at further distances.

\begin{figure*}[ht!]
\begin{center}
 \includegraphics[width=14.5cm]{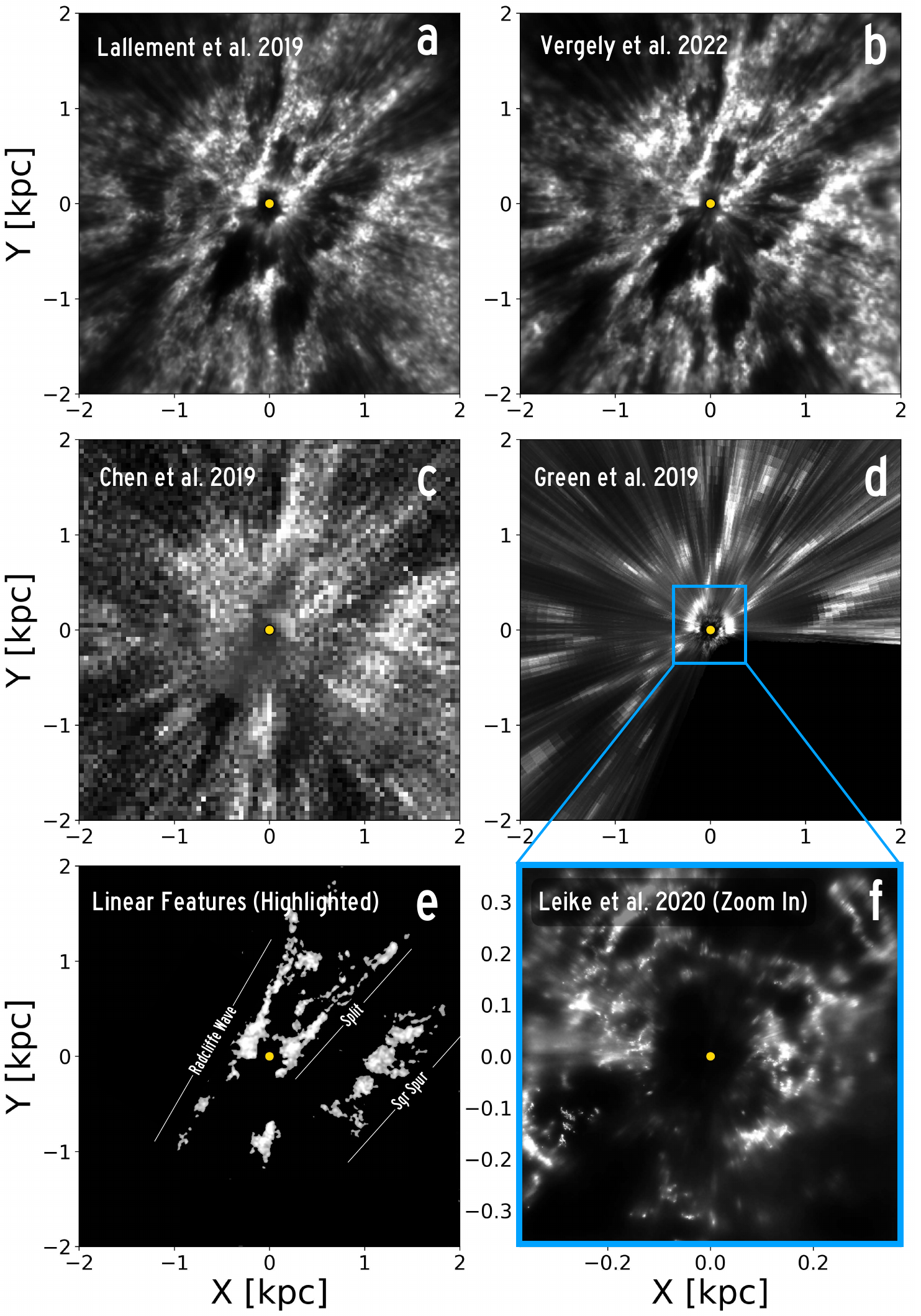}
 \caption{\small A bird's-eye view of four 3D dust maps in the \textit{Gaia} era out to 2 kpc and beyond, taken from (a) \citet{Lallement2019} (b) \citet{Vergely2022}, (c) \citet{Chen2018} and (d) \citet{Green2019}. In all cases, the extinction density is integrated between $z = \pm 300$ pc and displayed in arbitrary units. The Sun (yellow dot) is shown at the center. Linear kiloparsec-scale elongated features are seen in all four dust maps and are highlighted in panel (e); these features correspond to sections of spiral arms and spurs, as discussed in \S\ref{section5}. In panel (f) fully resolved cloud structure at $\approx$ 1 pc resolution is seen in the zoomed-in view of the nearest 400 pc \citep{Leike2020}, enabling new constraints on the internal structure of star-forming regions (see \S\ref{duststructure} and Figure \ref{fig:cloudstructure}).}
 \label{fig:dustmap}
 \end{center}
\end{figure*}

\subsubsection{\textbf{ The Gross Topological Distribution of the Interstellar Medium within 2 kpc of the Sun}} \label{grossdust}

Figure \ref{fig:dustmap} shows a heliocentric, bird’s-eye comparison of four 3D dust maps that provide the most complete coverage of the solar neighborhood out to at least 2 kpc, taken from \citet{Lallement2019,Vergely2022,Green2019,Chen2018}. We also include a zoom-in of the \citet{Leike2020} 3D dust map (out to 400 pc only; see \S\ref{duststructure}). Primarily built using a combination of photometry and \textit{Gaia} astrometry, these 3D dust maps probe the gross topological distribution of interstellar material on large scales and over a range of densities, spanning the atomic to the molecular medium. Several other 3D dust maps are also available in the \textit{Gaia} era but are not shown in Figure \ref{fig:dustmap}, either because they have very limited latitude coverage \citep{Hottier2020-pz}; are built on the first \textit{Gaia} data release \citep{Rezaei_Kh2018}; cover less than one-quarter of the sky \citep{Guo2021}; or have been recommended to be used with caution due to known artifacts in the 3D dust reconstruction \citep{Leike2022}.

As shown in Figure \ref{fig:dustmap}a, \citet{Lallement2019} combines 2MASS photometry with \textit{Gaia} DR2 astrometry using a Bayesian hierarchical inversion technique to estimate the distance and extinction to 27 million stars, constructing a 3D dust map which is complete out to 3 kpc from the Sun. The typical distance resolution of the map varies as a function of stellar density, achieving $\approx 25-50$ pc resolution in the nearest $\rm 1-2 \; kpc$, with declining resolution (up to $\approx 500$ pc) out to 3 kpc. In Figure \ref{fig:dustmap}b, we show the 3D dust map from \citet{Vergely2022}, which improves upon the 3D dust map from \citet{Lallement2019}. \citet{Vergely2022} leverages the same hierarchical inversion technique as \citet{Lallement2019}, but does so with a new astrophotometric catalog for 35 million stars with updated \textit{Gaia} EDR3 parallaxes \citep[see][]{Lallement2022}, as well as 6 million additional stars with spectroscopic constraints on their stellar properties. The result is a map with improved distance resolution and higher dynamic range. In Figure \ref{fig:dustmap}c, \citet{Chen2018} combines \textit{Gaia}, 2MASS, and WISE photometry with \textit{Gaia} DR2 astrometry to estimate the reddening to 56 million stars using a machine learning algorithm, constructing a map out to 6 kpc with full coverage of the Galactic plane over $|b| < 10^\circ$. The \citet{Chen2018} map has a significantly lower distance resolution than \citet{Lallement2019} ($\approx$ a few hundred parsecs) but a higher angular resolution, matching the \textit{Planck} resolution of 6' over much of the sky. Finally, presented in Figure \ref{fig:dustmap}d, \citet{Green2019} combine Pan-STARRS1 and 2MASS photometry with \textit{Gaia} DR2 astrometry for one billion stars in a Bayesian framework to produce a 3D dust map complete to 63 kpc over three-quarters of the sky. Like the \citet{Lallement2019} map, the nominal resolution of the Green et al. map varies as a function of distance, achieving $\approx 10$ pc resolution at 100 pc, $\approx 50$ pc resolution at 500 pc, and $\approx 200$ pc resolution at $\approx 2$ kpc. \citet{Green2019} also matches the 6’ angular resolution of \citet{Chen2018}.

The visible differences in the distance resolution of the 3D dust maps seen in Figure \ref{fig:dustmap} are primarily a combination of two factors: the stellar density per unit volume and the degree of spatial regularization imposed between neighborhood voxels. By spatial regularization, we mean that the 3D distribution of dust is correlated on multiple scales, and maps that exploit correlations between neighboring voxels can achieve higher resolution than maps that infer the dust distribution independently voxel-by-voxel. For example, even though the \citet{Chen2018} map uses a larger number of total stars in their reconstruction, the density of stars per unit volume is smaller than the \citet{Lallement2019} map, rendering a lower resolution reconstruction. And even though the \citet{Green2019} map includes many more stars per unit volume than \citet{Lallement2019} in the nearest 2 kpc, \citet{Lallement2019} imposes a larger degree of spatial correlation between neighboring voxels, rendering a higher resolution map despite the smaller stellar density per unit volume. \citet{Chen2018} imposes no correlation between neighboring voxels; \citet{Green2019} imposes correlations primarily on scales of $\approx 1-2$ pc; and \citet{Lallement2019, Vergely2022} adopt several different correlation scales depending on a variety of factors, including both the stellar density and stellar distance and extinction uncertainties.

Nevertheless, while all maps vary in distance (and angular) resolution, their defining features remain the same, despite employing different methodologies and photometric data. Across all maps, the dust density distribution appears to be dominated by elongated, linear features with similar orientations when viewed top-down (highlighted in Figure \ref{fig:dustmap}e). 
We will defer most discussion of large-scale structure and dynamics to §\ref{section5}, but note here that 3D dust mapping in the \textit{Gaia} era has ushered in a fundamentally new view of our solar neighborhood in a Galactic context. Maps such as those shown in Figure \ref{fig:dustmap} are not only redefining the structure of existing spiral features like the Local Arm \citep[Radcliffe Wave;][]{Alves2020}, but also revealing previously unknown spur/interarm structure \citep[the Split;][]{Lallement2019}. 3D dust maps are also revealing the structure of shells and superbubbles \citep[e.g. the Local Bubble, the Perseus-Taurus Superbubble, and the Orion-Eridanus superbubble;][]{Bialy2021, Joubaud2019, Pelgrims2020}, which manifest as cavities in the 3D dust density. The relationship between bubbles in 3D dust maps and newly formed stars in the solar neighborhood will be discussed further in \S \ref{triggeredsf}.

\begin{figure*}[ht!]
 \includegraphics[width=17cm]{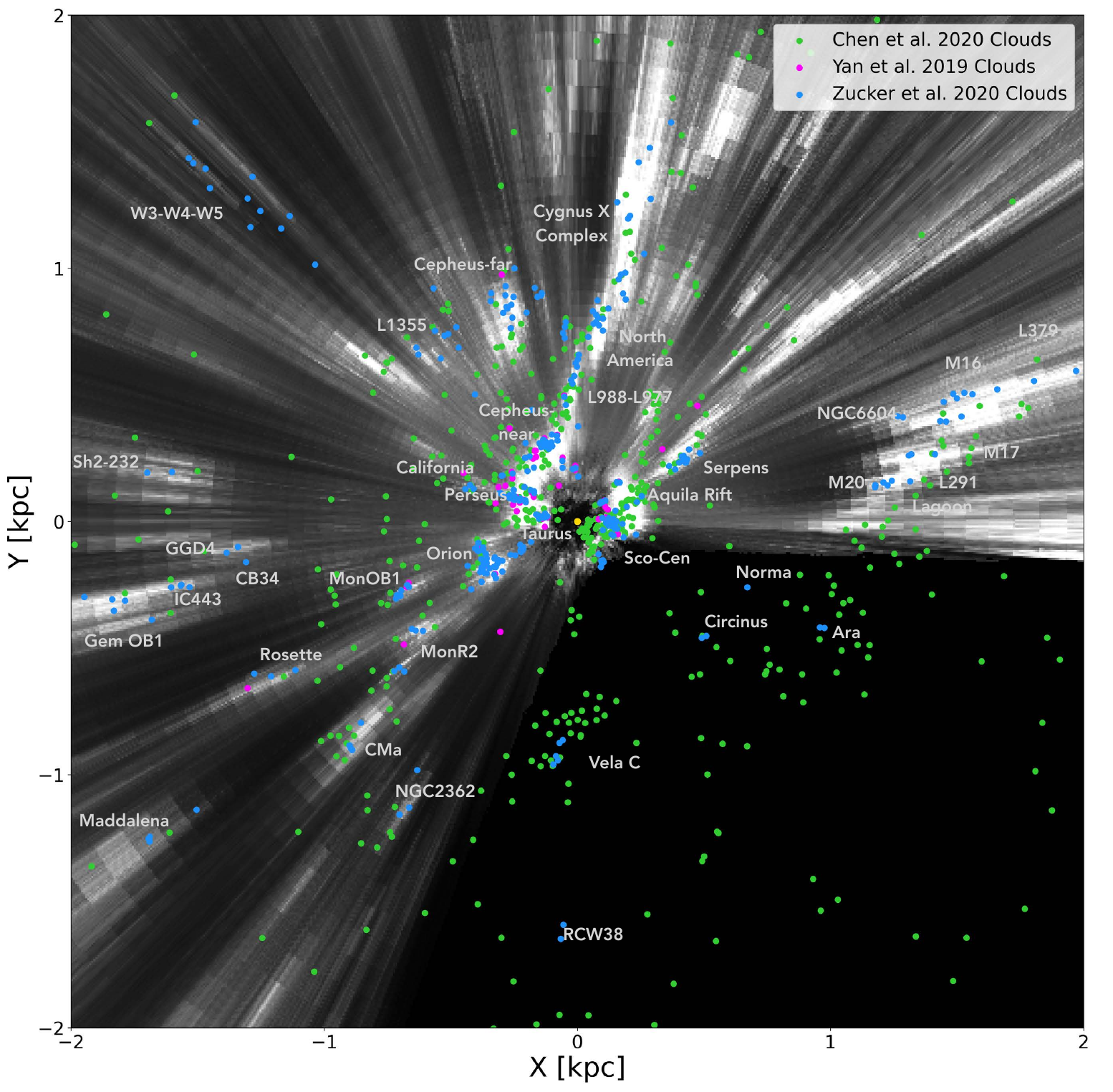}
 \caption{\small A bird's-eye view of the distribution of molecular clouds within 2 kpc from the same, taken from \citet{Zucker2020} (blue), \citet{Yan2019a} (magenta) and \citet{Chen2020b} (green). By optimizing traditional 3D dust mapping techniques (shown in Figure \ref{fig:dustmap}), distances to hundreds of molecular clouds in the solar neighborhood have been constrained with $\approx$ 5\% uncertainty, enabling new constraints on the global distribution of dense, star-forming gas. In the background grayscale, we show the 3D dust map from \citet{Green2019} (also shown in Figure \ref{fig:dustmap}d). An interactive version of this figure is available \href{https://faun.rc.fas.harvard.edu/czucker/Paper\_Figures/PPVII/3D\_Cloud\_Catalogs.html}{\textbf{\fbox{here}}}.}
 \label{fig:cloudcatalogs}
\end{figure*}

\subsubsection{\textbf{ Accurate Distances to Nearby Molecular Clouds based on 3D Dust Mapping}} \label{targeteddust}

The 3D dust maps shown in Figure \ref{fig:dustmap} provide the broadest coverage of the interstellar medium over a range of densities and at relatively high angular resolution. However, complementary 3D dust mapping techniques have been developed in the \textit{Gaia} era, sacrificing angular resolution for increased distance resolution towards the molecular star-forming interstellar medium. By targeting entire star-forming regions (or large subsets of star-forming regions), these cloud-focused methods achieve about $2\times$ better distance resolution ($\approx$ 5\% distance uncertainty) than 3D dust maps that take a more broad-brush approach, without sacrificing the ability to probe structure out to 2 kpc and beyond. 

Because these methods can be applied to any region with moderate amounts of dust and measurable stellar properties, they have ushered in a new era of homogeneous, accurate distance estimates to dozens of molecular clouds in the solar vicinity\index{Molecular cloud complexes}. While these 3D dust mapping techniques have been tailored to the denser interstellar medium, they are still largely probing the material around cores ($A_V$ $\lesssim$ 10 mag) — rather than the cores themselves — because of the requirement that stars must be visible behind the dust. Thus, while tailored 3D dust mapping techniques can provide the most accurate distances to a complete sample of nearby molecular clouds --- even those not actively forming stars --- they must be combined with complementary techniques (e.g. VLBI, \S\ref{vlbi}) to obtain distances to regions hosting the most deeply embedded protostars\index{protostars}. Many of these molecular clouds do indeed have complementary constraints on their distances based on VLBI and \textit{Gaia} astrometry of their young stellar populations, and we will discuss the good agreement between the two approaches in \S \ref{agreement}. 

In Figure \ref{fig:cloudcatalogs}, we overlay three uniform catalogs of accurate distances to local molecular clouds \citep[taken from][]{Zucker2020, Yan2019b, Chen2020a} on top of the \citet{Green2019} 3D dust map first presented in Figure \ref{fig:dustmap}. Broad agreement is seen between the two approaches, with the kiloparsec-scale elongated dust features first shown in the bird's-eye view in Figure \ref{fig:dustmap} clearly co-spatial with the molecular cloud complexes.

The Zucker et al. catalog (plotted in blue in Figure \ref{fig:cloudcatalogs}), combines optical and infrared photometry with \textit{Gaia} DR2 astrometry in a Bayesian framework to target sixty molecular clouds selected from the Handbook of Star-Forming Regions \citep{Reipurth2008a,Reipurth2008b}. The typical distance uncertainty of the \citet{Zucker2020} catalog is 5\%, and several distance samples are taken across each cloud, placing initial constraints on distance gradients and multiple components \citep{Zucker2019}. Alongside the \citet{Zucker2020} catalog, we overlay the distance catalog from \citet{Chen2020a} (green points), who apply a hierarchical structure identification algorithm to their 3D dust map \citep{Chen2018} to identify molecular clouds, before refitting those clouds distances’ with an optimized dust model to achieve distance uncertainties of $\approx$ 5\%. Finally, the \citet{Yan2019a} catalog is shown in magenta in Figure \ref{fig:cloudcatalogs}. Unlike Zucker et al. and Chen et al., Yan et al. use \textit{Gaia}’s own stellar distance and extinction estimates \citep{Andrae2018}, targeting a sample of molecular clouds at high galactic latitudes to achieve the same 5\% distance uncertainty. To validate the fidelity of typical distance uncertainties for all three catalogs, \citet{Chen2020a} calculate the dispersion of differences between their distance estimates and those from \citet{Yan2019a} and \citet{Zucker2019} for the same molecular clouds, finding a value of 5.6\%. Other catalogs not shown provide additional coverage of the solar neighborhood with similar distance uncertainties \citep{Yan2021b, Yan2021a, Yan2019b,Yan2020, Sun2021}.

There are a few limiting factors for achieving better than $5\%$ distance uncertainty over large swaths of molecular clouds out to many kiloparsecs. First, there is a limitation on how finely techniques can bracket clouds between unreddened foreground stars and reddened background stars, depending on both the total stellar density of stars along a cloud's line of sight and the signal-to-noise ratio of the parallax measurements per star. These cloud catalogs also assume a very simple slab-like model for the cloud's reddening at some distance, neglecting complexities in the cloud's 3D structure. And finally, there is also some systematic error in stellar modeling --- necessary to simultaneously constrain the distance and extinction of stars based on photometry --- which can cause systematic shifts in cloud distances. While systematic errors in stellar modeling will remain an issue at the $1-2\%$ level for the foreseeable future, continual improvements in \textit{Gaia} astrometry with future data releases, as well as more sophisticated models for cloud structure along the line of sight, should enable marginal gains in distance uncertainty in the coming years. 

Ultimately, despite the current 5\% error floor, these uniform catalogs of molecular cloud distances have enabled a wealth of star formation research\index{Giant molecular clouds!star formation}, from a physical characterization of cold clumps and cores extracted from the \textit{Planck} \citep{Yi2021}, JCMT \citep{Rumble2021} and \textit{Herschel} Gould Belt Surveys \citep{Fiorellino2020, Pezzuto2021}, to new constraints on the star formation scaling relations \citep{Pokhrel2020,Spilker2021} and a better understanding of the magnetic field structure around nearby shells and superbubbles \citep{Doi2021, Joubaud2019}.

\subsubsection{\textbf{The Internal Structure of Nearby Molecular Clouds}} \label{duststructure}
The 3D dust mapping results presented in \S\ref{grossdust} and \S\ref{targeteddust} represent significant gains in our understanding of the global distribution of molecular clouds. With typical distance uncertainties decreasing by a factor of a few compared to the pre-\textit{Gaia} era, we can now confidently associate particular clouds with larger-scale Galactic features, as well as derive cloud properties with fewer systematic biases. The 5\% distance uncertainty also enables some constraints on the structure of clouds along the line of sight, providing initial evidence for distance gradients or multiple components among a significant fraction of local clouds \citep{Zucker2019}. However, to fully resolve the internal structures of clouds (e.g., thickness) requires distance uncertainties of the order of 1\%, which even in the \textit{Gaia} era has been difficult to obtain, except in very rare instances where VLBI measurements are available (\S\ref{vlbi}).

Recently, \citet{Leike2020} presented a new 3D dust map of the solar neighborhood that constrains the structure of the interstellar medium at $\approx$ 2 pc resolution (1 pc grid) out to a distance of $\approx$ 400 pc from the Sun. Unlike many previous approaches, which suffer from a “fingers of god” effect --- owing to the angular resolution always being finer than the distance resolution --- Leike et al. infer the dust density directly in 3D ($x$, $y$, $z$) Cartesian space by combining Metric Variational Gaussian Inference with Gaussian Processes within the framework of Information Field Theory \citep{Arras2019,Enslin2019,Enslin2011}. Specifically, leveraging distance and extinction estimates for 5 million stars in the StarHorse catalog \citep{Anders2019} with excellent astrometric constraints from \textit{Gaia} DR2, Leike et al. reconstruct both the 3D dust density and its spatial correlation power spectrum by modeling the dust as a log-normal Gaussian process. 

The \citet{Leike2020} resolution limit of 2 pc was deduced based on their reconstruction of the spatial correlation power spectrum; however, there is also likely an additional small (but unknown) systematic error stemming from uncertainties in the stellar modeling of photometry used in the creation of the Starhorse catalog \citep{Anders2019}. However, these systematic errors should preferentially affect the absolute distances of the clouds rather than insights into their 3D structure. The high resolution of the \citet{Leike2020} map will be further validated in \S \ref{agreement}, where we will show strong agreement between the 3D dust maps of Taurus and Orion and VLBI measurements of their deeply embedded protostars \citep{Galli2019, Reid2019}.

In Figure \ref{fig:dustmap}f, we present the \citet{Leike2020} dust map alongside extant lower-resolution 3D maps, showing the elongated kpc-scale dust features in Figures \ref{fig:dustmap}a-\ref{fig:dustmap}d being decomposed into their constituent cloud complexes. This 800 pc $\times$ 800 pc bird's-eye view includes about a dozen fully resolved molecular clouds, including Perseus, Taurus, Cepheus, Orion, Corona Australis, and the Scorpius-Centaurus (Sco-Cen) association (Lupus, Ophiuchus, Chamaeleon, Pipe, and Musca). Figure \ref{fig:cloudstructure} shows zoom-ins of a subsample of these clouds, displaying the gas density in Heliocentric Galactic Cartesian space. Due to the map’s infancy, widespread characterization of the 3D structure of clouds has only just begun, but we highlight a few recent results by \citet{Zucker2021}, as well as similar studies carried out by \citet{Rezaei_Kh2020, Rezaei_Kh2022} and \citet{Dharmawardena2022a,Dharmawardena2022b}\index{Molecular cloud structure}.

Analyzing the dozen famous star-forming regions available in the Leike et al. footprint, \citet{Zucker2021} finds that local molecular clouds appear filamentary, even at lower gas densities. To characterize this filamentary structure, Zucker et al. apply the \texttt{FilFinder} algorithm \citep{Koch2015} to “skeletonize” the clouds in 3D volume density space and extract their “spines”, equivalent to a one-pixel-wide representation of the cloud’s 3D topology (shown in blue in the zoom-in panels of Figure \ref{fig:cloudstructure}). Zucker et al. determine an average radial volume density profile around each cloud’s 3D spine. Besides constraining the depth of each cloud, Zucker et al. find that the clouds’ radial profiles are well-described by a two-component Gaussian function, consistent with clouds having a narrow, higher-density inner layer and a broader, lower-density outer layer. Zucker et al. hypothesize that the transition between these two components could coincide with either a chemical transition (e.g., from HI to $\rm H_2$ gas) or a thermal transition (e.g., from the Unstable Neutral Medium to Cold Neutral Medium). 

Beyond constraining the thicknesses of clouds, Zucker et al. also project the clouds’ 3D spines back on the plane of the sky to obtain cloud distances with $\approx$ 1 pc uncertainty. Notable results include tenuous connections existing between the Chamaeleon molecular cloud and the Musca dark cloud, as well as between the Ophiuchus molecular cloud and the Pipe nebula; Taurus comprising two sheet-like layers; Corona Australis exhibiting a coherent, 70 pc long tail manifesting at lower gas densities; and prominent distance gradients/depth effects across Ophiuchus (35 pc), Perseus (22 pc), and Orion A (54 pc).

The prominent distance gradient across Orion A is consistent with earlier results from \citet{Rezaei_Kh2020} based on 3D dust mapping, as well as YSO-based results presented in \S \ref{stellar_distances}. Specifically, \citet{Rezaei_Kh2020} leverage \textit{Gaia} DR2 astrometry to constrain 3D stellar positions and WISE and 2MASS photometry to estimate extinction using the Rayleigh-Jeans Colour Excess method \citep[RJCE][]{Majewski2011} \citep[see also][]{Rezaei2018_Orion}. Modeling the set of stellar distances and extinctions toward Orion A using a Gaussian Process-based approach, Rezaei et al. construct a 3D map of Orion A at 10 pc resolution, finding it to be highly filamentary with a tail extending over 100 pc in distance. Rezaei et al. also detect a bubble-like structure in front of the cloud. This same dust shell is present in the Leike et al. map, which \citet{Swiggum2021} argue might be caused by stellar feedback stemming from two stellar groups near the core of the Orion complex (OBP-Near and Briceno-1). Mapping the 3D magnetic field morphology of the Orion region using both Faraday rotation measurements and polarized dust emission from \textit{Planck}, \citet{Tahani2022} find that 3D dust shells like these play a leading role in shaping the arc-shaped morphology of the magnetic field in the region \citep[see also e.g.][]{Tahani2019}.

Using updated astrometry from \textit{Gaia} EDR3, \citet{Rezaei_Kh2022} also apply their technique to the California Molecular Cloud, which has long been known to resemble Orion A on the plane of the sky based on similarities in their overall shapes, extents, and integrated extinctions. However, Orion A has a significantly higher star formation efficiency. Unlike Orion A, \citet{Rezaei_Kh2022} argue that California appears more sheet-like than filamentary and lacks the dense substructure that manifests across the Orion A cloud --- the combination of which may explain why California has a much lower star formation rate.

The importance of true 3D views on understanding the effect of cloud geometry (e.g. filamentary versus sheetlike) and feedback on molecular cloud properties is also well-demonstrated in \citet{Dharmawardena2022b}, who use their publicly available package \texttt{Dustribution} to model the 3D dust density in volume cutouts towards sixteen molecular cloud complexes \citep[see also][]{Dharmawardena2022a}, including the more distant regions of Cygnus, Carina, Rosette, and Canis Major. Applying the \texttt{astrodendro} package \citep{Robitaille2019} to extract molecular cloud boundaries and understand the larger cloud environments, \citet{Dharmawardena2022b} also find California to be much more massive and sheet-like than the Orion A complex. And in addition to confirming the existence of 3D cavities shaping the Orion cloud complex, \citet{Dharmawardena2022b} also find evidence of bubbles towards California, Chamaeleon, Rosette, and Canis Major. In most cases, the underlying source of these cavities is unknown but provides targeted opportunities for follow-up studies of stellar populations that might have produced stellar or supernova feedback in the recent past.\index{Superbubble shells}  

Building on the maps presented by Leike et al., Rezaei et al., and Dharmawardena et al., analysis of the internal structure of molecular clouds will only continue to improve in the \textit{Gaia} era, as more precise astrometric and photometric data are incorporated into existing pipelines --- rendering higher dynamic range maps --- and more creative techniques are applied to extract and characterize the structure present in extant maps. Moreover, the combination of the broad spatial coverage provided by the 3D dust maps in Figure \ref{fig:dustmap} and the higher-resolution, targeted views of molecular cloud structure discussed here will constitute powerful new datasets for understanding the small-scale physics of star formation within the broader Galactic environment.

\subsection{\textbf{Stellar-Based Distances to Nearby Star-Forming Regions}} \label{stellar_distances}

\S\ref{3ddust} has shown that 3D dust mapping can map out distances across wide swaths of clouds and their environs (e.g., including non-star-forming gas). However, \textit{Gaia} and VLBI astrometry provide the most precise distance determinations toward young stars and the compact high-gas-density regions that host them. Here we review distances to nearby star-forming regions based on stellar parallax measurements toward young stars, showing that the highest dynamic range reconstructions of individual star-forming regions’ distances and topologies come from combining these parallax measurements with 3D dust mapping.

\subsubsection{\textbf{Gaia-based Distances to Young Stellar Objects}}

Alongside the rise of 3D dust mapping, the exquisite astrometric precision of the \textit{Gaia} mission has enabled a robust inventory of optically visible young stellar objects — largely Class II and Class III protostars\index{Class II Sources}\index{Class II Sources}, with ages of a few to several million years. Because these protostars have not traveled far from their birth sites due to their youth, modeling the astrometry for young stellar populations across the clouds enables precise constraints on their distances, structure, and star formation histories. Determining distances to star-forming regions based on their young stellar populations has a long history, dating decades back to the Hipparcos mission \citep[see e.g.,][]{Bertout1999,Lombardi2008, Alves2007, Neuhauser1998}. The rise of \textit{Gaia} has not only increased the census of young stars in the solar neighborhood with astrometric distances from hundreds to tens of thousands, but it has also revealed much fainter, older, and more distributed populations than previously considered. 

In the \textit{Gaia} era, we start by highlighting these \textit{Gaia} young stellar results on a region-by-region basis toward the Taurus, Perseus, and Orion A molecular clouds, which were chosen due to their young ages ($\lesssim 10$ Myr) and because their natal molecular cloud environments remain intact. We then briefly remark upon similar case studies taking place for additional regions, before concluding by comparing distances and properties of local clouds based on a combination of young stars and 3D dust. We defer discussions of the more global distribution of young stellar associations (including stellar associations with ages $> 10$ Myr; cf. Sco-Cen) -- to \S\ref{section4} \citep{Kerr2021, Zari2018}.

\bigskip

\textbf{Taurus Molecular Cloud}\index{Taurus Molecular Cloud complex}: The Taurus region is one of the closest archetypes of a structured molecular cloud complex where a young population is present. The census of the stellar population compiled in previous studies \citep{Esplin2017, Joncour2017, Kenyon2008, Luhman2017} was revised by \citet{Luhman2018} after the publication of the \textit{Gaia} DR2 catalog. \citet{Luhman2018} manually divided the Taurus population into four subgroups which were labeled by color (blue, red, green, and cyan) based on their proper motions, parallaxes, and photometry. The blue and red populations overlap in the central portions of the complex hosting most of the Taurus star-forming clouds, while the green and cyan subgroups refer to the stars in L1558 and L1517/L1544, respectively. The existence of multiple populations in Taurus was also investigated by \citet{Roccatagliata2020} in a more recent study using Gaussian mixture models. In that study, the authors identified six populations (labeled Taurus A, B, C, D, E, and F), which are only partially consistent with the four subgroups given by \citet{Luhman2018}. Three of them (Taurus A, B, and E) are located in the central portions of the complex and roughly correspond to the red and blue populations in \citet{Luhman2018}. Taurus C and F host most of the stars in the northern and southern molecular clouds, respectively, while Taurus D is a more sparse population that contains stars from all four subgroups given in \citet{Luhman2018} and is potentially composed of multiple substructures. 

\citet{Galli2019} performed a hierarchical clustering analysis of the Taurus sample based on the HMAC algorithm \citet{Li2007} and identified 21 sub-populations (i.e., clusters). The clustering analysis with HMAC using the 5D astrometric phase-space allowed the detection of various substructures of the Taurus complex that were not resolved in other studies, leading to a more detailed picture of the structure of the region. Galli et al. confirmed the important depth effects of the region and showed that the distance to individual clouds ranges from 128.5 $\pm$ 1.6 pc (B 215) to 198.1 $\pm$ 2.5 pc (L1558) in good agreement with previous findings based on VLBI parallaxes \citep{Galli2018} and 3D dust (see \S\ref{duststructure}). Also applying a clustering algorithm to a census of young stars in Taurus with \textit{Gaia} EDR3 data, \citet{Krolikowski2021} identify seventeen groups, largely over the distance range $\approx 110 - 160$ pc including both clustered and distributed populations. Analyzing the ages of the groups, Krolikwoski et al. find evidence of at least two epochs of star formation (the oldest up to 15 Myr) and an increasing trend of cluster age with distance, suggesting that star formation in Taurus may be triggered from behind the cloud\index{Triggered star formation}. Evidence for this substantially older distributed Taurus population is also found in \citet{Liu2021,Kraus2012,Zhang2018} but their relationship to the younger clustered population in Taurus remains under debate \citep{Luhman2018, Luhman2022_taurus}.

\begin{figure*}[ht!]
 \includegraphics[width=17cm]{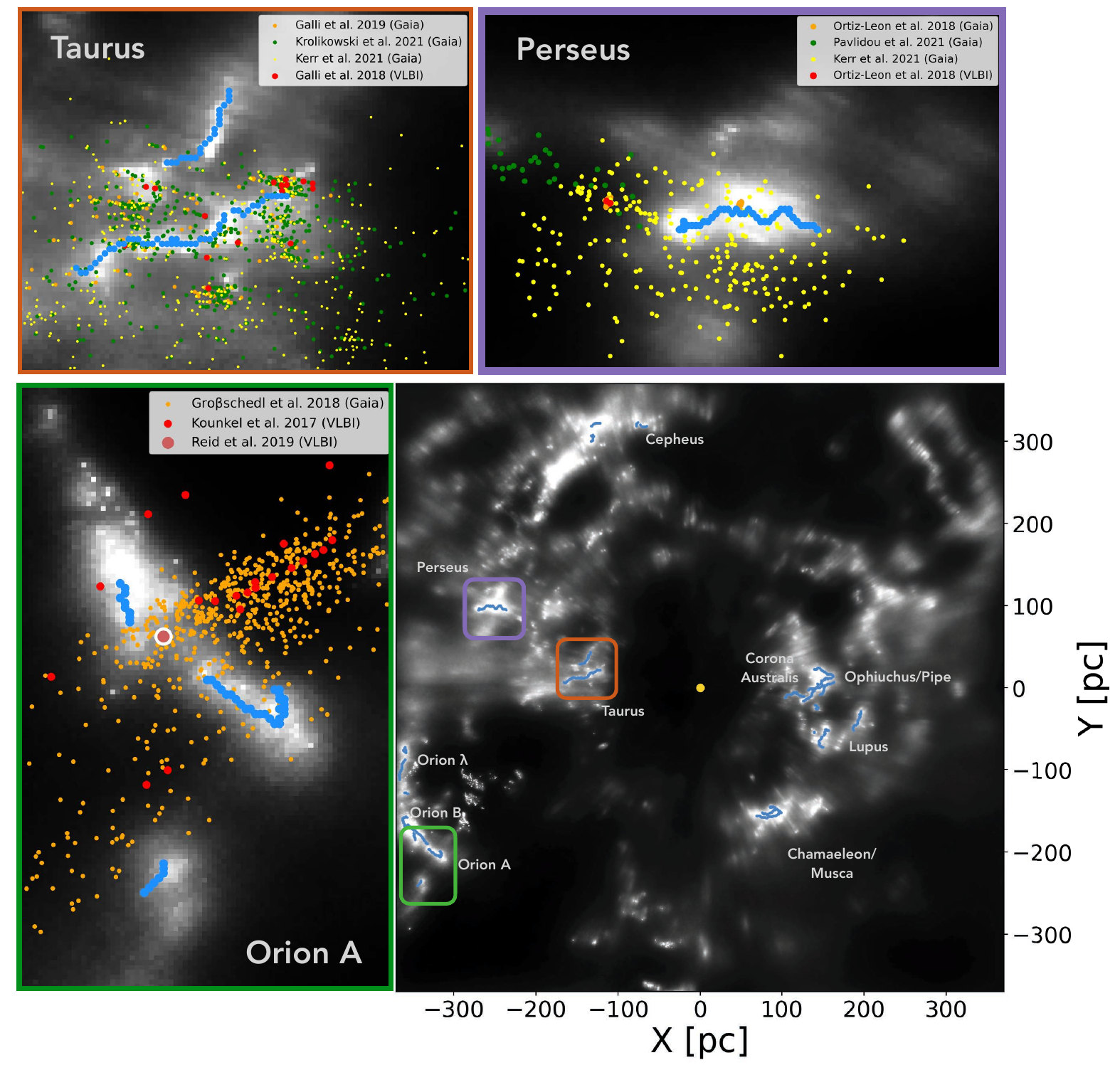}
 \caption{\small A top-down view of the solar neighborhood within 400 pc of the Sun. Background grayscale shows the integrated dust distribution from \citet{Leike2020}. Blue “skeletons” show the dense spines of nearby molecular clouds, labeled by name \citep{Zucker2021}. In the zoom-ins, we show detailed views of the cloud topology for the Taurus, Perseus, and Orion A molecular clouds, with stellar catalogs obtained from both \textit{Gaia} and VLBI overlaid on the 3D dust distribution, as stated in the legends. 3D interactive versions of the zoom-in panels are available \href{https://faun.rc.fas.harvard.edu/czucker/Paper\_Figures/PPVII/Taurus\_Cutout.html}{\textbf{\fbox{here}}} for Taurus, \href{https://faun.rc.fas.harvard.edu/czucker/Paper_Figures/PPVII/Perseus_Cutout.html}{\textbf{\fbox{here}}}for Perseus, and \href{https://faun.rc.fas.harvard.edu/czucker/Paper\_Figures/PPVII/OrionA\_Cutout.html}{\textbf{\fbox{here}}} for Orion A.} 
 \label{fig:cloudstructure}
\end{figure*}

A top-down view of the young stellar populations in the Taurus Molecular Cloud taken from \citet{Galli2019, Kerr2021, Krolikowski2021} and overlaid on the 3D dust map of \citet{Leike2020} is shown in one of the zoom-in panels of Figure \ref{fig:cloudstructure}. Overall, the emerging picture of the Taurus Molecular Cloud points to two prominent sheet-like structures corresponding to two distinct stellar populations separated by $\rm \approx 20-30$ pc in distance. The sheet-like morphology, coupled with evidence for multiple stellar populations of different ages, points to sequential star formation in action over at least the past $\approx 10$ Myr. Taurus has recently been discovered to lie wedged between two superbubbles --- the Local Bubble \citep{Pelgrims2020} and the Perseus-Taurus Supershell \citep{Bialy2021,Doi2021} --- which may likely explain both its sheet-like morphology and its sequential star formation activity \citep{Zucker2022}

\bigskip

\textbf{Perseus Molecular Cloud}: Like the Taurus Molecular Cloud, the Perseus Molecular Cloud is an ideal laboratory for studying low-mass star formation, hosting a chain of famous dark clouds, including B5, IC 348, B1, NGC 1333, L1455, L1448, and L1451. A prominent $\rm 10 \; km\;s^{-1}$ velocity gradient is observed from west (L1451) to east (B5) of the complex. Most of the young stars reside in the IC 348 and NGC 1333 clusters with ages between $\rm \approx 1-3 \; Myr$, which together host a few hundred members detected with \textit{Gaia}. Cross-matching the young stellar census of \citet{Luhman2016} toward IC 348 and NGC 1333 with \textit{Gaia} DR2 data, \citet{Ortiz-Leon2018} carry out a systematic analysis of the 3D distribution of young stars in the Perseus complex. 

The most notable result is a prominent 30 pc distance gradient between NGC 1333 ($\rm d = 293 \; pc$) and IC 348 ($\rm d = 320 \; pc$), consistent with its large velocity gradient from east to west. Building on the work of \citet{Ortiz-Leon2018}, \citet{Pezzuto2021} looks beyond IC 348 and NGC 1333 to characterize structure across all sub-regions in Perseus. Cross-matching young stars originally identified in \textit{Herschel} data with \textit{Gaia}, Pezzuto et al. identify thirty sources across L1451, L1448, NGC 1333, B1, IC 348, and B5. Pezzuto et al. confirm the same prominent distance gradient from west to east, spanning about 40 pc in total ($\rm d = 282 - 325 \; pc$). 

Finally, looking at a much larger area than previous studies, \citet{Pavlidou2021} apply several astrometric and photometric cuts to the \textit{Gaia} data to identify approximately 6,000 young star candidates near the Perseus cloud. Apart from the prominent NGC 1333 and IC 348 clusters, Pavlidou et al. identify five new young stellar populations: four populations are located off the main Perseus cloud with ages of $3-5$ Myr, while the last population lies on cloud with an age $<$ 1 Myr and a similar distance to NGC 1333. Like the Krolikowski et al. results in Taurus, Pavlidou et al. note a general sequence of star formation in the Perseus cloud, with stellar populations at farther distances (closer to the Galactic plane) having larger ages compared to the nearer, on-cloud populations at higher latitudes. Validating a new sample of \textit{Gaia}-identified young stars with LAMOST spectroscopy --- including some of the \citet{Pavlidou2021} groups --- \citet{Wang2022} also finds evidence of more distributed stellar population around the Perseus Molecular Cloud. 

A top-down of the young stellar populations in the Perseus Molecular Cloud \citep{Kerr2021, Ortiz-Leon2018, Pavlidou2021} overlaid on the 3D dust map of \citet{Leike2020} is also shown in one of the zoom-in panels of Figure \ref{fig:cloudstructure}. The cumulative evidence points to a significant distance gradient ($\approx 30$ pc), significant velocity gradient ($\rm \approx 10 \; km \; s^{-1}$), and significant age gradient ($\approx 5$ Myr) gradient across the Perseus Molecular Cloud. Given that Perseus lies on the edge of the recently discovered Perseus-Taurus Supershell \citep{Bialy2021,Doi2021}, it is likely that the structural, dynamical, and temporal evolution of the region has also been heavily affected by superbubble feedback in its recent past \citep[see also][]{Wang2022}, as it has also been in the Taurus Molecular Cloud (see further discussion on the Perseus-Taurus Supershell in \S\ref{section4}). 

\bigskip

\textbf{Orion A Molecular Cloud}: The Orion A molecular cloud is the most active site of star formation within 500 pc, having produced a few thousand young stellar objects over the past few megayears \citep{Megeath2012, Grossschedl2019}. Unlike Taurus and Perseus, which host low-mass stars, Orion A is also the most famous and nearest site of high-mass star formation, making it the focus of several investigations in the \textit{Gaia} era to determine its 3D shape and orientation. 

One of the first studies, by \citet{Grosschedl2018}, used over 700 \textit{Spitzer}-selected YSOs with available \textit{Gaia} DR2 data to determine that Orion A is actually composed of two components: a denser, actively star-forming bent head ($\rm d = 400 \; pc$) and a more diffuse 75-pc-long tail with less active star formation (extending to $\rm d = 470 \; pc$). Großschedl et al. argue that the bent head and tail structure seen in 3D could be due to either a cloud-cloud collision or stellar feedback. A subsequent study by \citet{Getman2019} confirms Orion A’s unique bent head-tail morphology and also suggests that a compressive shock driven by OB feedback could be responsible. The concept of ongoing stellar and supernova feedback shaping cloud structure is a consistent theme throughout the literature on Orion A, with complementary studies by \citet{Kounkel2020_supernova} and \citet{Grosschedl2021} both pointing to a major supernova event occurring about 6-7 Myr ago \citep[see also][]{Kounkel2022_Orion}. Full characterization of such feedback events requires information not just on the 3D positions of young stars, but also on their 3D velocities. New systematic studies seeking to expand upon the known stellar populations of Orion — uncovering new spatially and kinematically distinct subgroups spanning a range of ages — will be critical for constructing a sequence of star formation over the past 20 Myr \citep{Chen2020, Jerabkova2019, Kos2021, Marton2019, Swiggum2021, Zari2019} and will be discussed further in \S \ref{triggeredsf}.

A top-down of the young stellar populations in the Orion A Molecular Cloud \citep{Grosschedl2018} overlaid on the 3D dust map of \citet{Leike2020} is shown in the final zoom-in panel of Figure \ref{fig:cloudstructure}. The combination of stellar and dust-based studies points to a cloud that is highly filamentary, whose structure has been shaped by energetic supernova feedback activity, and whose young stars are a product of large-scale triggering. 

\bigskip

\textbf{Beyond Taurus, Perseus, and Orion}: The results presented for Taurus, Perseus, and Orion A represent just a sample of the breadth and depth of 3D structure work being performed for other star-forming regions across the solar neighborhood. We refer the reader to similar case studies for the Cepheus \citep{Szilagyi2021, Saha2021, Klutsch2020, Szegedi-Elek2019, Kerr2022, Sharma2022}, Chamaeleon \citep{Galli2021, Kubiak2021, Roccatagliata2018, Sacco2017}, Lupus \citep{Galli2020-Lupus, Luhman2020, Melton2020}, Corona Australis \citep{Galli2020-CorAus, Esplin2022}, Ophiuchus \citep{Canovas2019, Esplin2020, Grasser2021, Gupta2022}, Serpens \citep{Herczeg2019, Zhou2022}, Cygnus \citep{Molina_Lera2021, Orellana2021, Rao2020, Comeron2020}, North American Nebula \citep{Kong2021, Kuhn2020}, Gum Nebula \citep{Yep2020}, Trifid Nebula \citep{Kalari2021,Kuhn2022}, NGC 2264 \citep{Flaccomio2022, Buckner2020}, Rosette \citep{Lim2021, Muzic2022}, Mon OB1 \citep{Lim2022_Mon}, Carina \citep{Goppl2022}, NGC 6334 \citep{Russeil2020}, IC 1396 \citep{Pelayo-Baltarrago2022, Silverberg2021, Aguilar2019}, and CMa OB1 \citep{Gregorio-Hetem2021, Santo-Silva2021} star-forming regions for more details.

\subsubsection{\textbf{VLBI Distances to Star-Forming Regions}} \label{vlbi}

While \textit{Gaia} provides accurate distances to Class II and Class III protostars, it is an optical instrument and cannot probe the densest star-forming regions due to dust extinction. To do so, one needs radio observations and recent VLBI surveys like the BeSSeL \citep{Reid2019, VERA_collaboration2020} and the Gould’s Belt Distances Survey (GOBELINS) \citep{Loinard2012} channel the strengths of \textit{Gaia} at much longer wavelengths\index{Interferometry}. At radio wavelengths, there are two types of sources bright enough to be detected with VLBI --- masers associated with high-mass star-forming regions, and compact radio emission caused by the gyration of electrons near magnetically active, lower-mass young stars\index{Masers}.

The BeSSeL survey has determined accurate distances to around 200 high-mass star-forming regions to date, including about thirty in the nearest 2 kpc, with typical uncertainties $<$ 10\%. The nearest major star-forming region with a BeSSeL measurement is associated with the Orion Nebula Cluster\index{Orion Nebula Cluster}. Similarly, the first release of the VERA astrometry catalog has constrained the distances to 99 masers with similar uncertainties, approximately half of which lie within 2 kpc. To complement both BeSSeL and VERA, the GOBELINS survey has targeted low-mass star-forming regions throughout the Taurus \citep{Galli2018}, Orion \citep{Kounkel2017}, Perseus \citep{Ortiz-Leon2018}, Serpens \citep{Ortiz-Leon2017b}, and Ophiuchus molecular clouds \citep{Ortiz-Leon2017a}, with distance uncertainties to individual clumps on the order of $\rm \approx 1-2 \; pc$. GOBELINS and BeSSeL data for the Taurus, Perseus, and Orion regions \citep{Galli2018, Kounkel2018, Ortiz-Leon2018, Reid2019} are shown alongside the \textit{Gaia}-visible YSOs in the Figure \ref{fig:cloudstructure} zoom-in panels and overlaid on the \citet{Leike2020} 3D dust map. 

\subsection{\textbf{Agreement between Dust, YSO, and Maser-based Distance Estimates}} \label{agreement}
\begin{figure}[h]
 \includegraphics[width=1.0\linewidth]{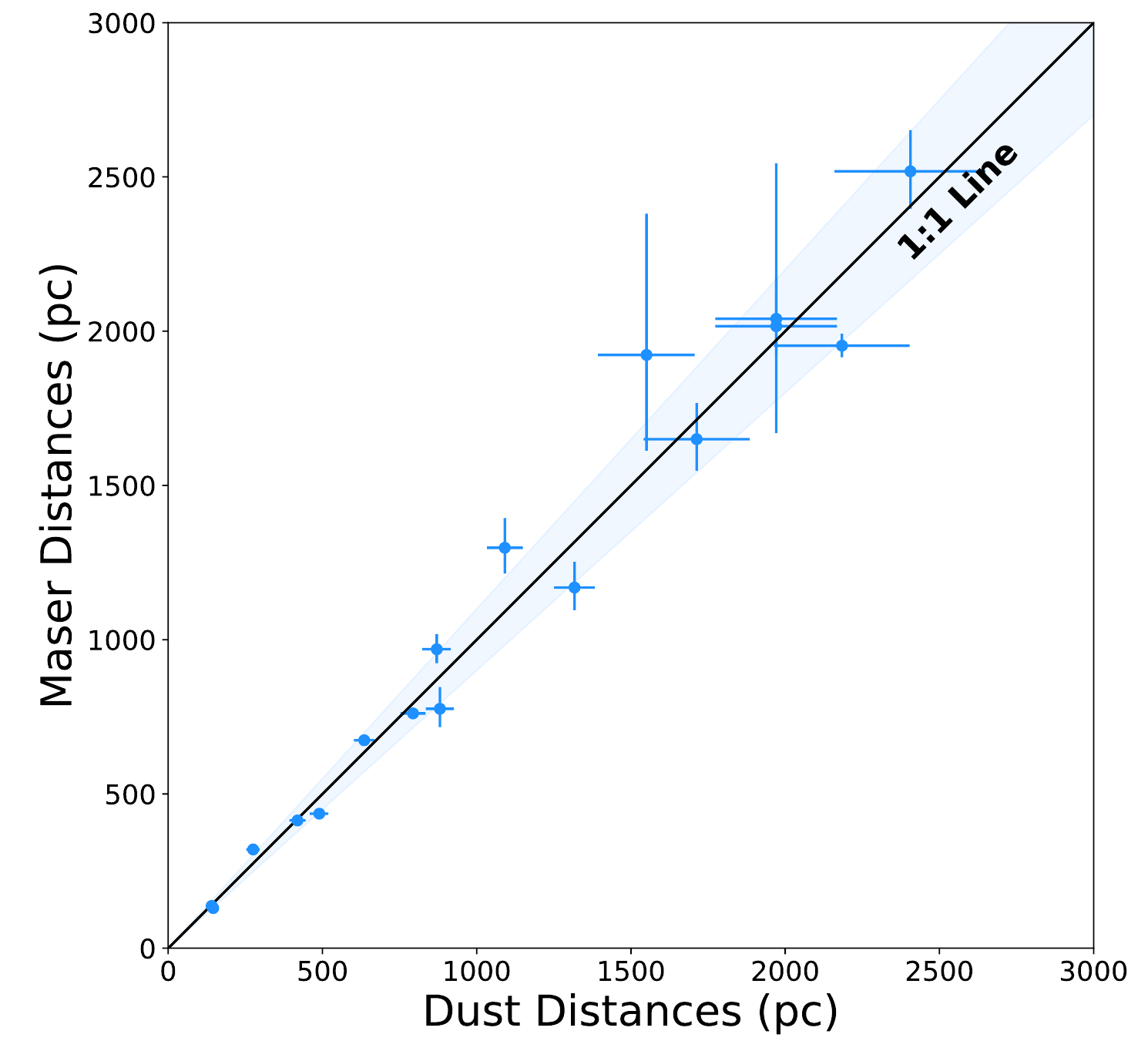}
 \caption{\small Distances derived from 3D Dust Mapping versus distances derived from VLBI toward masers and low-mass stars. Good agreement is seen out to 2 kpc, with typical scatter between the two approaches $<$ 10\%. Credit: Zucker et al., A\&A, 633, A51, 2020, reproduced with permission \copyright ESO.}
 \label{fig:stars_vs_dust}
\end{figure}

While 3D dust mapping, \textit{Gaia} parallax measurements toward YSOs, and VLBI parallax measurements toward masers and radio-active low-mass stars offer three independent constraints on cloud distances and their internal structure, broad agreement should be expected between the three techniques. In Figure \ref{fig:stars_vs_dust}, we show how good the agreement is between dust-based distances and VLBI-based distances (from BeSSeL and GOBELINS)\index{Masers}. The typical scatter is $<$ 10\% across large swaths of the solar neighborhood including nearby star-forming regions like Taurus \citep{Galli2018}, Perseus \citep{Ortiz-Leon2018}, and Orion \citep{Kounkel2018, Reid2019} as well as more distant star-forming regions like W3 and NGC2362 \citep{Reid2014, Reid2019}.

In the zoom-in panels of Figure \ref{fig:cloudstructure}, we again highlight detailed comparisons between the structure of the Taurus, Perseus, and Orion A star-forming regions defined by 3D dust and that defined by VLBI and \textit{Gaia} constraints on young stars. 

Toward the Taurus star-forming region, we see evidence of the depth effects and the multiple distance components (one at $\rm d \approx 130 \; pc$ and one at $\rm d \approx 150 \; pc$) in both the 3D dust and in the young stellar populations constrained by \textit{Gaia} and VLBI. However, we also see that the optically visible, and older, protostars detected with \textit{Gaia} \citep{Galli2019, Kerr2021, Krolikowski2021} show more dispersions around the 3D dust \citep[blue skeletons; taken from][]{Zucker2021} than the GOBELINS radio sources \citep{Galli2018}, which should have higher distance accuracy and probe earlier stages of star formation. The agreement (to within $\rm \approx 2-3 \; pc$) between the VLBI GOBELINS sources and the 3D dust distribution validates the $\approx 1\%$ distance uncertainty of the \citet{Leike2020} map for the Taurus star-forming region. 

Toward the Perseus Molecular Cloud, there is excellent agreement between the 3D dust and the YSOs toward the younger NGC 1333 population ($\approx$ 1 Myr). However, the older IC 348 population \citep[$\approx$ 3 Myr;][]{Luhman2003} detected with both \textit{Gaia} and VLBI ($\rm d = 321 \pm 10 \; pc$) appears about 20 pc beyond the dust cloud. While agreement is still good (only discrepant by $\approx$ 5\% in distance), it is likely that some of this variation in the spatial structure of gas and young stars is physical, especially since \citet{Zucker2018} obtain a similar 3D dust distance as constrained in \citet{Leike2020}. We expect young stars to evolve out of their parental clouds as they age, be ejected as runaways from the cloud due to e.g. interactions with other members \citep{Oh2016}, or for the cloud itself to be dispersed by stellar feedback \citep{Walch2012}, underlying the potential of a combined \textit{Gaia}, VLBI, and 3D dust approach to constrain the physics of how stars leave home. While beyond the scope of this review, the underlying mechanisms for cloud and star dispersal can be better constrained by contrasting the distance offsets with the age offsets for different subgroups, given the known velocity dispersions of the gas and young stars --- a prospect made even more possible by the advent of new radial velocity data from \textit{Gaia} DR3. 

Finally, toward the Orion A cloud, there is excellent agreement between the sole BeSSeL measurement \citep{Reid2019} within the nearest $\approx$400 pc (toward the Orion Nebula Cluster) and the 3D dust, with the maser measurement aligning with the ``spine” of the cloud to within a few parsecs in this top-down view. Since masers by definition have to be associated with very dense and high-mass-star-forming gas, this agreement again underlies the outstanding accuracy of the 3D dust \citep{Leike2020} in the \textit{Gaia} era. The rest of the optically visible stellar population is offset from the 3D dust cloud by about $\rm 10-15 \; pc$ closer in distance. Unlike in \citet{Rezaei_Kh2020}, evidence of the connected ``tail” detected in YSO-based \textit{Gaia} studies is not seen in the 3D dust from \citet{Leike2020} (manifesting as two discrete distance components), and follow-up work will be needed to shed light on this discrepancy.

\section{\textbf{CLOUD MOTIONS AND KINEMATICS IN 3D}} \label{section3}

In this section, we review the 6D results (3D space and 3D velocity) in young stellar groups, emphasizing the most embedded populations in star-forming regions, as they are most likely to trace any motion inherited from their parental cloud. 

Using Taurus, Lupus, Chamaeleon, and Corona Australis as examples, we discuss (i) how distance variations along the line of sight in many regions (first introduced in \S\ref{section2}) map to variation in their 3D space motions (ii) detection of substructures with different kinematic properties and their relative motions (iii) dynamical effects such as expansion, contraction, and rotation that have become apparent in some star-forming regions in the \textit{Gaia} era and (iv) comparison of the kinematics of the young stars with the underlying gaseous clouds based on the CO emission of the molecular gas. 

Given the infancy of this field, our goal in this section is to lay the foundation for what will become even more possible in the upcoming years with future data releases of \textit{Gaia} and other large-scale public surveys based on what we have already learned from current data. 

\subsection{\textbf{Taurus 3D Motion}}\label{taurusmotion}

As introduced in \S\ref{3ddust}, Taurus is both the nearest major molecular cloud to the Sun and also the prototypical laboratory for understanding low-mass star formation. The Taurus young stellar populations are not randomly distributed in the sky but are clustered in clumps which map to the morphology of the molecular clouds and filaments constructed from CO surveys and 2D extinction maps in the past \citep{Cambresy1999, Dame2001, Dobashi2005, Goldsmith2008, Ungerechts1987}.
 
The 3D space motions of the clouds reveal the existence of internal motions within the complex that justify the velocity dispersion of several $\rm km \; s^{-1}$ reported in past studies \citep{Bertout2006, Jones1979}. For example, \citet{Galli2019} computed the relative motion of $\rm 3.2 \pm 0.5 \; km \; s^{-1}$ between the northernmost and southernmost clouds, but demonstrated that the relative motion among the various clouds can reach up to about $\rm 5 \; km \; s^{-1}$. The observed velocity dispersion in the Taurus region is the combination of random and organized motions. Galli et al. demonstrated that the internal motions in the radial direction of the complex are dominated by random motions, implying that the complex is neither expanding nor contracting. They found evidence of a potential rotation of the complex \citep{Rivera2015} that requires confirmation and further investigation with more precise data. The large-scale survey of the Taurus molecular clouds in $\rm ^{12}CO$ and $\rm ^{13}CO$ performed by \citet{Goldsmith2008} was also used in that study to compare the kinematics of the stars and the molecular gas in this star-forming region. The good agreement between the radial velocity of the stars and the velocity of the gas measured along the line of sight at the position of the stars is confirmed by the non-significant mean difference between the velocity of the two components of $\rm 0.04 \pm 0.12 \; km \; s^{-1}$ with an RMS of $\rm 0.63 \; km \; s^{-1}$. 

\begin{figure*}[h!]
 \centering
 \includegraphics[width=0.9\linewidth]{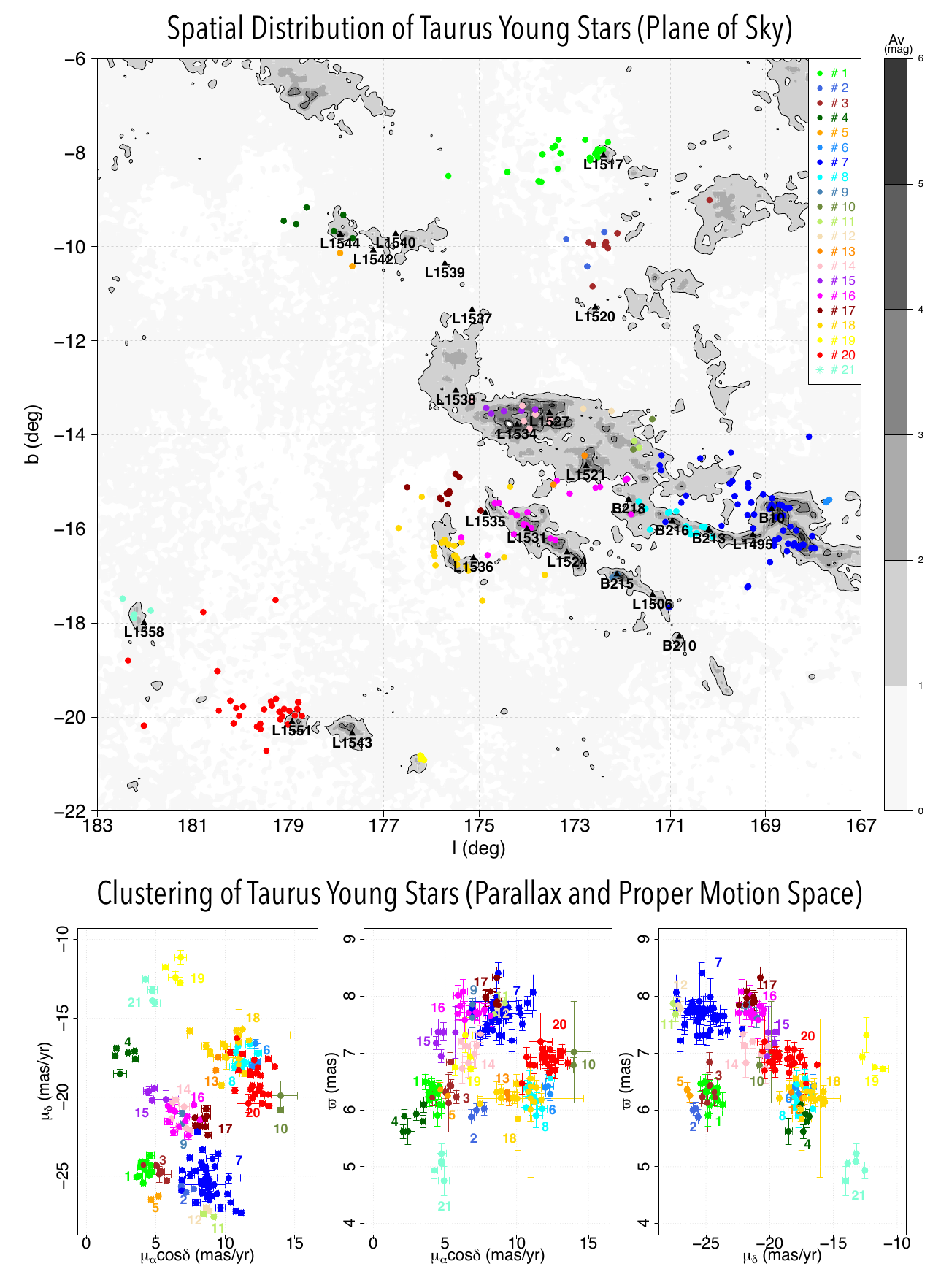}
 \caption{\small Upper panel: Spatial distribution of Taurus stars overlaid on the extinction maps from \citet{Dobashi2005} in Galactic coordinates. The different colors denote the Taurus subgroups identified by \citet{Galli2019}. Lower panels: Clustering of Taurus stars in the space of proper motions and parallaxes \citep[see][]{Galli2019}. Credit: Galli et al., A\&A, 630, A137, 2019, reproduced with permission \copyright ESO.}
 \label{fig:taurus}
\end{figure*}

This result confirms that the kinematic properties of the stars and gas are tightly coupled, implying that the observed velocity field of the underlying molecular cloud complex maps to the velocity of the stars. One interesting finding discussed by Galli et al. is the existence of two families of molecular clouds in the Taurus complex that exhibit different velocity vectors and can be recovered directly from the clustering analysis with the HMAC algorithm, potentially suggesting different formation episodes for the various clouds in the region (see Figure \ref{fig:taurus}).

This evidence for multiple star formation episodes in the complex is further strengthened by the more recent results of \citet{Krolikowski2021}, who characterize the spatial distribution, dynamics, and ages of seventeen distinct subgroups in the larger Taurus complex using \textit{Gaia} EDR3 and radial velocities drawn from the literature. As introduced in \S \ref{stellar_distances}, Krolikowski et al. determine that Taurus has experienced at least two epochs of star formation over the past 15 Myr, featuring both clustered and distributed stellar populations. Moreover, Krolikowski et al. find that the distributed populations have larger velocity dispersions than the clustered population, consistent with the expectations of Larson’s Second Relation, whereby the velocity dispersion of the clusters scales with the stellar separation amongst the cluster members, providing new insight into how the properties of the young stellar populations are tied to the initial conditions of the molecular gas. 

The current picture of the star formation history in Taurus is still incomplete as it requires further investigation in the context of a larger, extended star formation event in the solar neighborhood and at least two epochs of star formation to explain the existence of the clustered and distributed subgroups throughout the complex \citep{Krolikowski2021}\index{Clustered star formation}. In addition, more studies will be necessary to refine the census of the stellar population based on the 3D space motion of the stars, enable a 6D picture of the structure of the complex, and unravel the (complicated) history of star formation in Taurus. Most current studies are still operating in a 5D space of parameters (3D positions and 2D tangential velocities) due to the lack of (precise) radial velocity measurements, which can cause projection effects and break up large subgroups in a clustering analysis. However, with the recent release of new radial velocity measurements in \textit{Gaia} DR3, studies utilizing the full 6D phase-space information of young stars should become more widespread. 

\subsection{\textbf{Lupus 3D Motion}}

The Lupus region hosts nine molecular clouds (labeled Lupus 1-9), which differ significantly in terms of their stellar population and star formation activity. These clouds are distributed at distances between $\rm 150-200 \; pc$, and are likely distributed over two layers \citep{Zucker2021}. Early studies in Lupus have mostly focused on the Lupus 1-4 molecular clouds containing most of the known candidate members in this region \citep{Comeron2008}. Recent studies using \textit{Gaia} data have also identified a few young stellar objects in Lupus 5 and 6 \citep{Manara2018, Melton2020, Teixeira2020}. The pre-\textit{Gaia} studies in the Lupus region based on trigonometric parallaxes from the Hipparcos catalog \citep{Bertout1999} and kinematic parallaxes inferred from the moving cluster method \citep{Galli2013,Makarov2007} reported on the existence of important depth effects in this region. 

This has recently changed with the release of \textit{Gaia} DR2 and now \textit{Gaia} DR3, which has allowed for a more accurate picture of the 6D structure of the Lupus complex. \citet{Galli2020-Lupus} showed that the Lupus 1-5 subgroups are roughly located at the same distance of about 160 pc. The distances of Lupus 1, 2, and 5 are less accurate because they were computed from a small sample of members, but the more robust results obtained for Lupus 3 ($\rm d = 158.9 \pm 0.7 \; pc$) and Lupus 4 ($\rm d = 160.2 \pm 0.9 \; pc$) confirm that their distances are consistent. Lupus 3 is the most extended subgroup, with distances to individual stars ranging from about 154 to 162 pc. The 3D space motion of the stars reveals that the Lupus subgroups are moving at a common speed, and the relative motion between the subgroups is not significant. Galli et al. also reported on the existence of a few more dispersed stars in Lupus (labeled as off-cloud population) which share similar kinematics and age of the stars projected towards the molecular clouds confirming that the various subgroups in Lupus (including on-cloud and off-cloud stars) are co-moving and therefore belong to the same association. 

Lupus is located on the sky between the Upper Scorpius (US) and Upper Centaurus Lupus (UCL) subgroups of the Sco-Cen association. The \textit{Gaia} studies in this region addressed many past uncertainties regarding membership in the Lupus clouds and Sco-Cen \citep{Damiani2019, Luhman2020, Teixeira2020}. Surprisingly, many of the historically known members of the Lupus region (with available astrometry in the \textit{Gaia} DR2 catalog) have been assigned to the Sco-Cen association or rejected as field stars. \citet{Damiani2019} revisited the stellar population of the Sco-Cen complex and identified 10,839 pre-main sequence stars as members belonging either to compact or diffuse populations. Most of the stars clustered in compact groups appear to be younger than the diffuse population, suggesting that star formation occurred in rapidly dispersed small groups\index{Clustered star formation}. The Lupus region is presented as one of these compact groups of the Sco-Cen complex in the study conducted by Damiani et al. The 3D spatial overlap of the Lupus clouds with the diffuse populations is inevitable, which explains the difficulties encountered in past studies to assign membership to the stars in this region. 

However, Lupus appears to be spatially distinct from other compact groups in the Sco-Cen complex. The closest substructure in Sco-Cen to the Lupus clouds is a compact group near V1062 Sco called UCL-1 \citep{Roser2018, Damiani2019} which is farther away from the Sun (d $\approx$ 175pc) according to Galli et al. The relative motion between Lupus and US and UCL respectively is ($\Delta U$, $\Delta V$, $\Delta W$) = (+2.4, -0.7, -0.4) $\pm$ (0.4, 0.2, 0.1) $\rm km \; s^{-1}$, and ($\Delta U$, $\Delta V$, $\Delta W$) = (+2.1, +2.4, -1.6) $\pm$ (0.4, 0.2, 0.1) $\rm km \; s^{-1}$ \citep{Galli2013, Wright2018}, where $U$, $V$, and $W$ are the motions along the Heliocentric X, Y, and Z directions. This implies that Lupus is moving with a relative bulk motion of 2.5 $\pm$ 0.4 km/s and 3.6 $\pm$ 0.3 $\rm km \; s^{-1}$ with respect to US and UCL, respectively. Thus, the 3D space motion of the Lupus clouds appears to be more consistent with US. Interestingly, the observed difference between the space motion of the Lupus clouds and Sco-Cen is comparable to the difference in spatial velocity reported among the Sco-Cen subgroups themselves. Therefore, Lupus can be considered another substructure of the Sco-Cen association, but the younger age suggests that it resulted from a more recent star formation episode.

\subsection{\textbf{Chamaeleon 3D Motion}}

Chamaeleon hosts the southernmost star-forming region in the solar neighborhood. It comprises three molecular clouds (Cha I, Cha II, and Cha III), but only two of them (Cha I and Cha II) have ongoing star formation activity. Cha I itself is composed of two subclusters that have been historically separated into the northern and southern subgroups based on their declinations \citep{Luhman2007, Luhman2008}.

The distance to the Chamaeleon complex has become a matter of debate over the past decades, with values ranging from 115 to 400 pc \citep{Fitzgerald1976, Franco1991, Grasdalen1975, Hughes1992}. \citet{Whittet1997} derived the distance estimates of $165 \pm 15$ pc for Cha I and $178 \pm 18$ pc for Cha II based on multiple distance indicators later confirmed from trigonometric parallaxes of the Hipparcos catalog \citep{Bertout1999}. One major step towards establishing the distance to Chamaeleon was done by \citet{Voirin2018} based on \textit{Gaia} DR1 data. In that study, the authors derived the distances of $\rm 179^{+11}_{-10} \; pc$, $\rm 181^{+11}_{-10} \; pc$, $\rm 199^{+12}_{-11} \; pc$ for Cha I, Cha II, and Cha III, respectively, placing Cha I about 20 pc further away and delivering the first distance determination of Cha III. However, the uncertainties of the resulting distances were largely dominated by the systematic errors in the \textit{Gaia} DR1 catalog, demanding further investigation. \citet{Roccatagliata2018} used \textit{Gaia} DR2 parallaxes to derive the distances of 192.7 $\pm$ 0.4 pc and 186.5 $\pm$ 0.7 pc for the northern and southern subgroups of the Cha I molecular cloud, respectively. More recently, \citet{Galli2021} derived the distances of 191.4 $\pm 0.8$ and 186.7 $\pm 1.0$ pc for the northern and southern subgroups of Cha I, respectively, and $\rm 197^{+1.0}_{-0.9} \; pc$ for Cha II also based on \textit{Gaia} DR2 data but using Bayesian inference and considering the spatial correlations of the stars in the analysis. The distances to individual stars obtained by Galli et al. range from about 184 to 194 pc, and from 194 to 202 pc in Cha I and Cha II, respectively. In that study, the authors also concluded that the Cha I and Cha II molecular clouds are roughly separated by about 23 pc in the 3D space of positions. 

The two subgroups of Cha I exhibit a mean difference in proper motions of about 1 $\rm mas \; yr^{-1}$, which translates into a difference of about 1 $\rm km \; s^{-1}$ in tangential velocity at the cloud's distance \citep{Galli2021}. In addition, \citet{Sacco2017} also reported that the mean difference in the radial velocity of the two subgroups is roughly 1 $\rm km \; s^{-1}$. So, the non-trivial question that arises is whether the observed differences in proper motion and radial velocity result from a different space motion of the two subgroups or simply projection effects. Galli et al. compared the space motion of the two subgroups in Cha I and showed that the relative motion between them is not significant when correlated errors that result from the transformation of parallaxes, proper motion, and radial velocities to 3D velocities are considered. Thus, the current picture suggests that the two subgroups in Cha I are located at different distances but exhibit consistent space motion within their uncertainties. The same conclusion holds for the observed difference in the 3D space motion between Cha I and Cha II, which are also not significant within the reported uncertainties. In particular, Galli et al. argue that the internal velocity dispersion of Cha II is not resolved because of the relatively large uncertainties in the radial velocities of the stars that are currently available in the literature and propagate directly to the 3D spatial velocities. Thus, the observed scatter in the spatial velocity of Cha II stars is more likely explained by measurement errors. 

Recently, \cite{Ratzenbock2022-mr} found that despite the Chamaeleon molecular clouds being at the periphery of the Sco-Cen complex, their 3D space motion is consistent with them belonging to the Sco-Cen association. The same conclusion can be drawn for the young moving groups $\epsilon$ Cha and $\eta$ Cha. As seen from the Sun, $\epsilon$ Cha lies in between the Chameleon clouds in projection only, as there is about 100 pc distance between the clouds (in the background) and the $\epsilon$ and $\eta$ Cha stellar associations (in the foreground).

The general connection between these stellar groups suggests that they belong to the same overall structure connecting the Ophiuchus clouds to the Sco-Cen complex that has been forming stars over the last $\approx$ 20 Myr and is part of the large-scale star-forming structure “the Split” \citep{Lallement2019}, extending a few kiloparsecs in length (see \S\ref{section5}). 

\begin{figure*}[h!]
\centering
 \includegraphics[width=14cm]{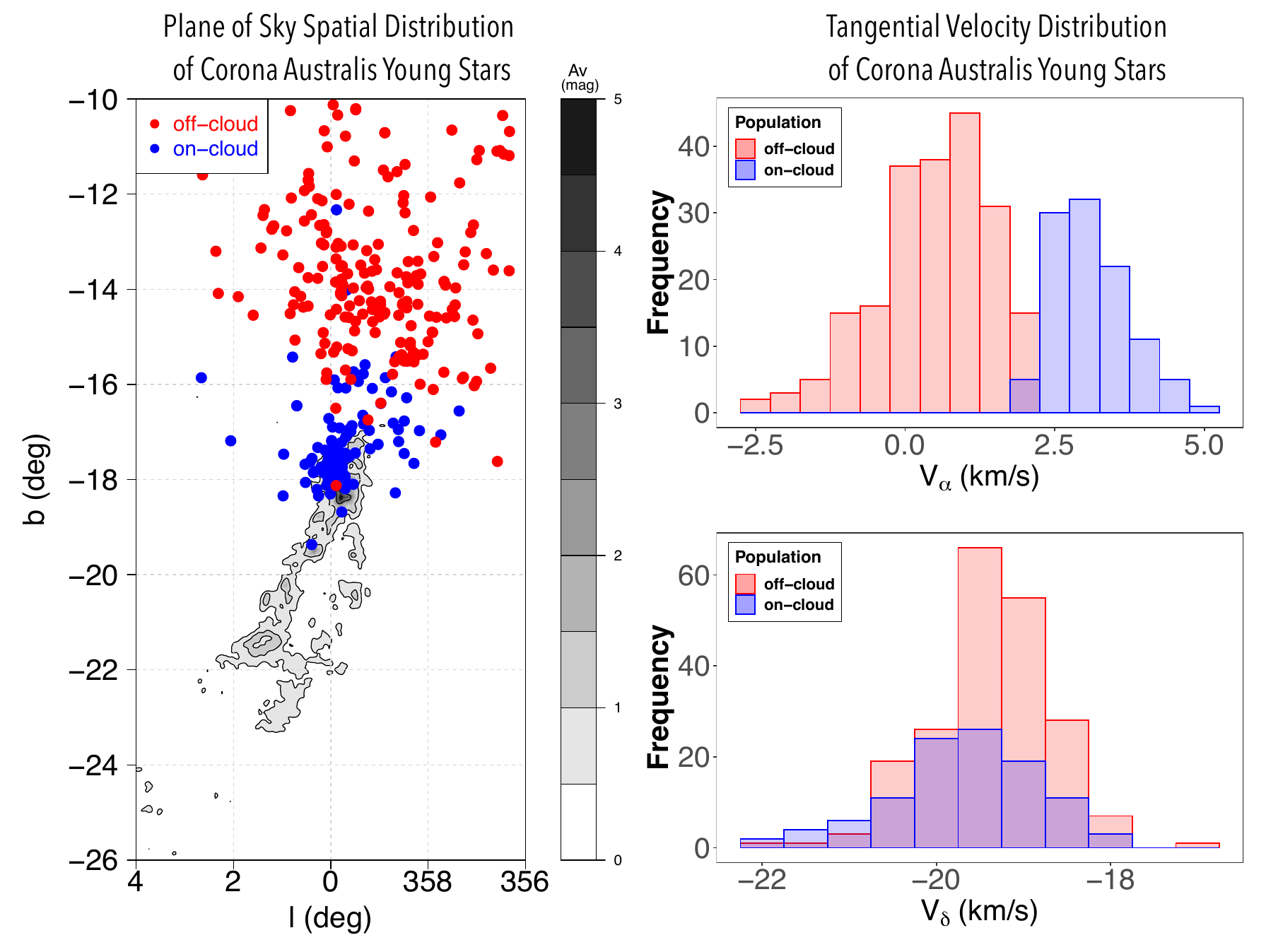}
 \caption{\small Left panel: Location of the Corona Australis stars identified by \citet{Galli2020-CorAus} overlaid on the extinction maps from \citet{Dobashi2005} in Galactic coordinates. Right panels: histogram of the tangential velocities for the two subgroups in Corona Australis \citep{Galli2020-CorAus}. Credit: Galli et al., A\&A, 634, A98, 2020, reproduced with permission \copyright ESO.}
 \label{fig:coraus}
\end{figure*}

\subsection{\textbf{Corona Australis 3D Motion}}

The star-forming head of the comet-like Corona Australis cloud is about $18^\circ$ below the Galactic plane and is relatively isolated from the other star-forming regions of the southern sky. Still, it is considered a part of the Sco-Cen complex \citep{Mamajek2000}. The canonical distance of about 130 pc commonly used in the literature for Corona Australis was based on the orbital motion of one star in the region, namely TY CrA, located at $\rm 129 \pm 11 \; pc$ \citep{Casey1998}. After the publication of the \textit{Gaia} catalog, \citet{Galli2020-CorAus} estimated the distance of $\rm 149.4 \pm 0.4 \; pc$ from \textit{Gaia} DR2 parallaxes with the individual distances to stars ranging from 134 to 169 pc), placing the entire star-forming region about 20 pc further away. 

The existence of a dispersed population of young stars around the dark clouds in Corona Australis was first pointed out by \citet{Neuhaeuser2000}. The authors identified 19 pre-main sequence stars aged about 10 Myr in this study. More recently, \citet{Gagne2018a} considered the \citet{Neuhaeuser2000} populations a separate stellar group in the solar neighborhood and named it Upper CrA. Galli et al., based on \textit{Gaia} DR2 data, revealed that the dispersed stars identified in pre-\textit{Gaia} studies belong to a much larger population of young stars, which also hosts the B2V star HR 7029 as a cluster member.

The census of the stellar population in Corona Australis obtained from pre-\textit{Gaia} studies lists 149 stars, including members and candidate members \citep{Neuhauser2008, Peterson2011}. However, many of these sources were not observed by the \textit{Gaia} satellite, while others exhibit proper motions and parallaxes that are not consistent with membership in this region. In a recent study, \citet{Galli2020-CorAus} revisited the census of the Corona Australis population, confirming 51 stars from the literature and identifying another 262 cluster members. Interestingly, the authors reported the existence of a more distributed ``off-cloud" population of young stars ($\approx$ 10 Myr) in the Galactic north of the dark clouds (see Figure \ref{fig:coraus}) containing twice as many as the historical members concentrated around the densest cores of the region. 

The existence of the two populations in Corona Australis is also clearly apparent from the distribution of proper motions and tangential velocities (see Figure \ref{fig:coraus}). The most discriminant astrometric feature in the \textit{Gaia} DR2 catalog between the two subgroups is the proper motion component in Right Ascension, which translates into a mean difference of about 2-3 $\rm km \; s^{-1}$ in the tangential velocity of the two subgroups. If we convert the stellar proper motions to the galactic reference system, then the proper motion component in Galactic Longitude becomes the most discriminant feature between the two subgroups. So, the different distribution of proper motions and 2D tangential velocities of the two subgroups is independent of the reference system that is used in the analysis.

The question that arises is whether the observed difference in the tangential velocities results from a different 3D space motion or because of projection effects stemming from different sky positions of the two populations. The lack of radial velocity measurements for most stars in the two populations prevents us from deciding between the two scenarios. \citet{James2006} derived the average radial velocity of $\rm -1.1 \pm 0.5 \; km \; s^{-1}$ for Corona Australis, but this result is based on a small sample of cluster members that is a mix of “on-cloud” and “off-cloud” stars. It is apparent that the tangential velocity is the dominant component in the spatial velocity of the stars in Corona Australis. Knowledge of the 3D space motion of the stars is also important to reconstruct the origin and local history of star formation. For example, \citet{Mamajek2001} suggested that Corona Australis stars formed near the UCL subgroup of the Sco-Cen association in the past and moved away to their current position. It is now possible to investigate this scenario by tracing the stellar positions back in time pending precise characterization of the 3D space motion of the stars in the two subgroups of Corona Australis.

Collectively, \textit{Gaia} has opened a new window for star formation studies based on the 3D space motion of the stars. It is now feasible to investigate the local history of star formation and young stellar clusters using the kinematic properties of the stars and the underlying gaseous clouds. One critical ingredient in our understanding of the history of star formation and feedback processes in star-forming regions is the age of the stars. Recent studies have demonstrated the possibility of deriving dynamical ages of young stellar groups by integrating the stellar orbits back in time using the 3D positions and 3D space motions of the stars \citep{MiretRoig2020,Miret-Roig2022, Miret-Roig2018}. The traceback technique is therefore a powerful tool to derive dynamical ages and reconstruct the evolution of the cluster in time that is not dependent on stellar models, complementing traditional approaches based on stellar isochrone fitting. Future studies combining the 3D positions, 3D space motions, and ages of the stars will allow us to produce a complete 7D picture of the closest starbursts to the Sun. We will introduce some of these earliest 7D studies near the Sun in \S \ref{section4}.

\bigskip
\noindent

\section{\textbf{YOUNG STELLAR STRUCTURES AND THE INTERSTELLAR MEDIUM}} \label{section4}

This section will focus on the distribution of young stellar populations, with a particular focus on their relationship to the gas structures from which they have evolved. Building on new constraints on 3D cloud motions from \S \ref{section3}, we will start by examining evidence for triggered star formation in the solar neighborhood. We will then highlight new studies on the closest young stars to the Sun, and in particular toward the Sco-Cen stellar association. We will conclude by looking into young stellar streams and cluster coronae as the link between embedded systems and the Galactic field.

\subsection{\textbf{Star Formation and Stellar Feedback: Wind-Blown Bubbles in the Nearest 500 pc}} \label{triggeredsf}

Positive stellar and supernova feedback triggers the creation of new stars by sweeping up dense shells of gas in a collect-and-collapse process, resulting in tenuous, hot cavities bounded by neutral shells of gas and dust \citep[see e.g.][]{Elmegreen2011}. High-resolution 3D dust maps made with \textit{Gaia} data have revealed a wealth of such nearly empty voids in the local Milky Way, many of which show evidence of star formation on their surface. The detection of these 3D cavities complement traditional integrated tracers of recent stellar and supernova feedback based on X-ray \citep[e.g.][]{Mayer2022}, $H\alpha$ \citep{Finkbeiner2003}, and $\rm ^{26}Al$ emission \citep{Diehl2008} associated with energetic stellar progenitors responsible for clearing their environments. Here, we discuss four feedback events producing interstellar cavities and likely giving rise to their associated young stellar structures: the Local Bubble \citep{Zucker2022}, the Orion Big Blast Event \citep{Grosschedl2021}, the Vela Shell \citep{Cantat-Gaudin2019b}, and the Perseus-Taurus Supershell \citep{Bialy2021,Doi2021}.\index{Triggered star formation}\index{Superbubble shells}  

\begin{figure*}[ht]
 \includegraphics[width=17cm]{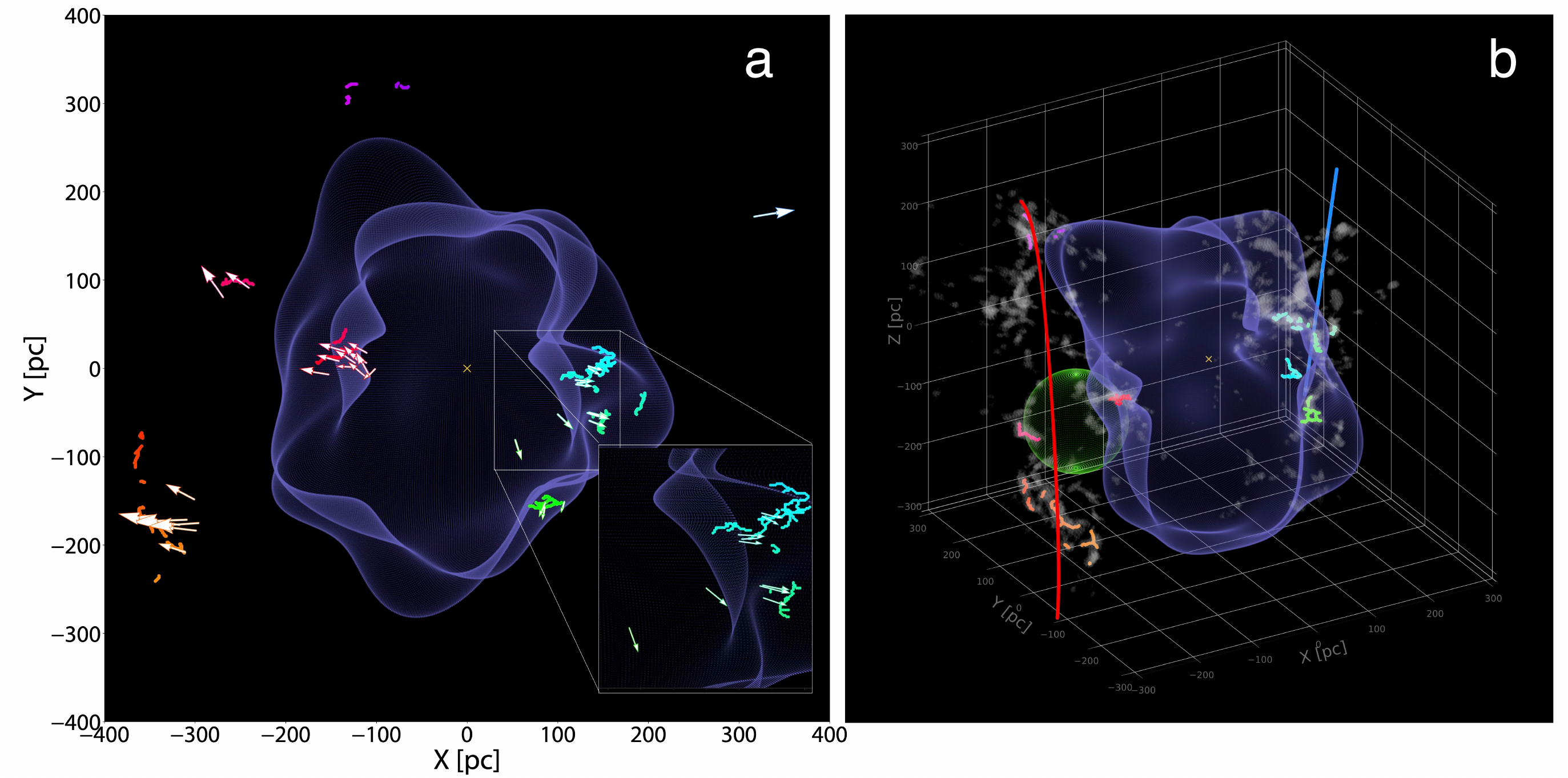}
 \caption{\small (a) Star formation on the surface of the Local Bubble. The purple surface shows the model for the Local Bubble from \citet{Pelgrims2020}. The squiggly colored lines show the positions of star-forming gas from \citet{Zucker2021}, along with their young stellar clusters, whose 3D positions and 3D motions are marked with the arrows. (b) In the right-hand panel, the Perseus-Taurus Supershell is shown \citep[green sphere;][]{Bialy2021, Doi2021} along with the Radcliffe Wave \citep[red linear trace;][]{Alves2020}, the Split \citep[blue linear trace;][]{Lallement2019}, and the 3D dust from \citet{Leike2020} (grey blobby shapes). Supernova explosions beginning 14 Myr ago in the Sco-Cen association triggered the formation of the expanding bubble and likely gave rise to a new generation of star-forming regions on its surface. An interactive version of this figure is available \href{https://static-content.springer.com/esm/art\%3A10.1038\%2Fs41586-021-04286-5/MediaObjects/41586\_2021\_4286\_MOESM2\_ESM.html}{\textbf{\fbox{here}}} and an animation showing the expansion of the Local Bubble over the past 14 Myr is available \href{https://static-content.springer.com/esm/art\%3A10.1038\%2Fs41586-021-04286-5/MediaObjects/41586\_2021\_4286\_MOESM3_ESM.html}{\textbf{\fbox{here}}}. Reproduced from \cite{Zucker2022}, \textit{Nature}, 601, 334.}
 \label{fig:local_bubble}
\end{figure*}

\subsubsection{\textbf{The Local Bubble }}
\label{LocalBubble}

The Local Bubble is an extensive region of lower-than-average interstellar density around the Sun, identified via a variety of techniques in the second half of the twentieth-century \citep{Lucke1978, Sanders1977}. In 2020, \citet{Pelgrims2020} used the map of \citet{Lallement2019} to create a three-dimensional model of the surface of the Local Bubble (see purple surface in Figure \ref{fig:local_bubble}). 

Displaying the models of 3D cloud structure from \citet{Zucker2021} (squiggly-colored lines in Figure \ref{fig:local_bubble}) in the context of the \citet{Pelgrims2020} model, it is immediately apparent that all nearby dense star-forming regions lie on the surface of the Local Bubble. \citet{Zucker2022} point out this coincidence and use \textit{Gaia} data on the 3D space motions to study the past trajectories of the young star clusters associated with each of the star-forming regions, finding clear evidence for a global expansion of the Local Bubble.

The expansion of the $\approx$320-pc-wide Local Bubble has long been suspected to be caused by past supernovae in the solar neighborhood \citep{Breitschwerdt2016, Fuchs2006, Maiz-Apellaniz2001}\index{Supernova}. Moreover, supernova bubbles have been responsible for sweeping up much of the dense gas that forms stars in galaxies as they expand and sometimes collide \citep{Inutsuka2015, McKee1977}. When finding clear evidence for an expanding gas shell with star formation on its surface, \citet{Zucker2022} modeled the Local Bubble expansion \citep{El-Badry2019} over time as driven by a group of supernovae originating in the Upper Centaurus Lupus (UCL)/Lower Centaurus Crux (LCC) clusters, about 14 million years ago. Further evidence for the supernova origin of the Local Bubble arises from the $\rm ^{60}Fe$ record; $\rm ^{60}Fe$ is a radioactive isotope produced predominantly in supernova explosions and shows two major influxes in the $\rm ^{60}Fe$ deposits detected in deep-sea sediment over the past 10 million years \citep{Wallner2021,Schulreich2018}. Precise ages of the young stellar clusters associated with the shell's surface provide a constraint on the evolution of the shell as a function of time. 

The time evolution of the modeled bubble’s expansion is extraordinarily consistent with all other nearby young star clusters that have formed on the surface of the Local Bubble as it expands over the past 14 Myr. The bubble, therefore, constitutes a prime example of sequential star formation\index{Sequential star formation}. The Bubble's expansion likely contributed to the rise of the Taurus, Corona Australis, Ophiuchus, Musca, Lupus, and Chamaeleon star-forming regions, many of which show evidence of spatial, kinematic, and age gradients in their young stellar populations as previously discussed in \S \ref{section2} and \S \ref{section3}.

\begin{figure*}[h!]
\centering
\includegraphics[width=14cm]{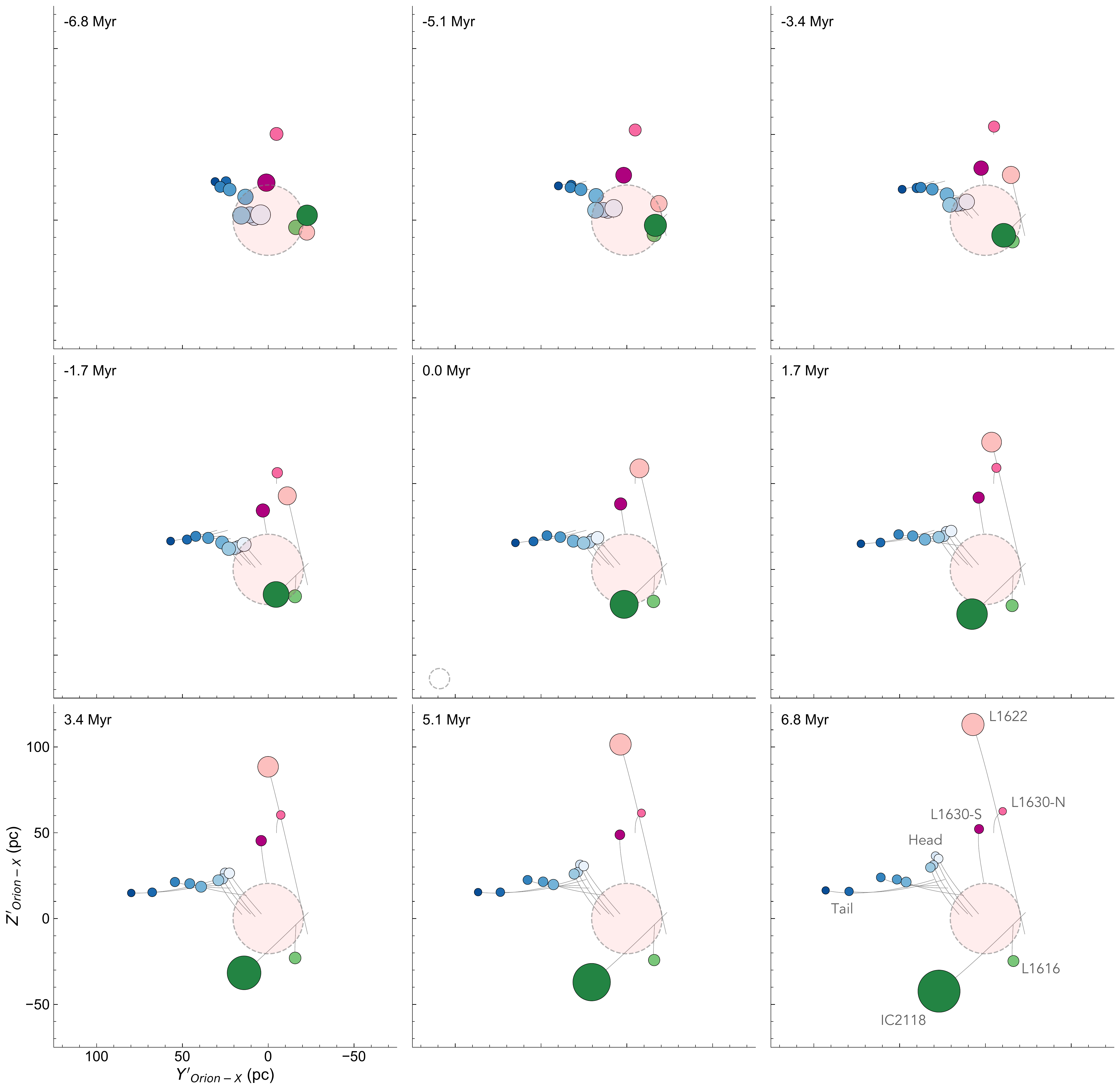}
\caption{\small Time-snapshots showing the relative motions of subregions in the Orion star-forming region from -6.8 to 6.8 Myr. This projection represents a side view of the Orion cloud complex and includes the massive Orion A and Orion B clouds. The coordinate system is centered on the potential source of feedback in the region, represented by the red disk with a gray dashed outline. The 14 Orion subregions of Orion A and B are shown as filled circles. The large-scale expansion can be seen as traced by several of the subregions, and points to a substantial feedback event (named the Orion Big Blast event, or Orion-BB event) that took place about 6 Myr ago. The observed expansion requires the energy of several supernovae. An interactive 3D version of this figure can be found \href{https://www.aanda.org/articles/aa/olm/2021/03/aa38913-20/aa38913-20.html}{\textbf{\fbox{here}}}. Credit: Gro{\ss}schedl et al., A\&A, 647, A91, 2021, reproduced with permission \copyright ESO.}
\label{fig:Orion-BB}
\end{figure*}

\subsubsection{\textbf{Orion Big Blast Event}}

As introduced in \S \ref{section2}, the morphology of the Orion complex has been shaped by supernova feedback over at least the past 10 Myr.
Recently, \cite{Grosschedl2021} used the parallaxes and proper motions of young stellar objects (YSOs) from \textit{Gaia} DR2 as a proxy for gas distance and proper motion and the gas radial velocities from archival CO data to compute the space motions of the different star-forming clouds in the Orion complex, including the large Orion A and Orion B clouds, and two outlying cometary clouds. These authors find that the different Orion subregions were closest about 6 Myr ago and are moving radially away from the same region in space (see Figure \ref{fig:Orion-BB}). This coherent 100-pc scale radial motion supports a scenario where the entire complex is reacting to a major feedback event that took place 6 Myr ago, which they name the Orion-BB (big blast) event. They argue that this event might be connected to the Orion-Eridanus superbubble. This event, which the authors tentatively associate with the stellar population of Orion X \citep{Chen2020} --- but could probably be linked to other clusters \citep[Briceño-1/25 Ori and the Orion Belt Population/OBP-B1;][]{Swiggum2021} --- shaped the distribution and kinematics of the gas we observe today. 

Building on the work of \citet{Grosschedl2021}, \citet{Foley2022} confirm the expansion is either stemming from the Orion X or OBP-B1 association. Using 3D dust maps \citep{Leike2020} to characterize the gas topology of the region, \citet{Foley2022} find that the OBP-B1 cluster resides in a dust cavity -- now termed OBP-B1-Cav. The cluster \textit{also} resides in the center of the prominent $H\alpha$ feature Barnard's Loop, and the molecular clouds Orion A, Orion B, and Orion \(\lambda\) reside in the shell of this cavity in 3D. The OBP-B1-Cav is also likely a sub-cavity of the larger Orion-Eridanus supershell \citep{Joubaud2019}, which has been known to be carved out by a combination of ionizing UV radiation, stellar winds, and a sequence of a supernova explosion. Thus, while it is unlikely that this Orion-BB event was the sole major feedback event in the region, its effects are measurable today on the 3D gas motion and structure, and, likely, at least some fraction of the stellar populations in Orion A, B and, $\lambda$ have been triggered by this ongoing feedback activity. 

The impact of feedback on the turbulence of gas in the Orion region is further supported by the work of \citet{Ha2021}, who compute the velocity structure function using 3D positions and 3D motions of young stars in the Orion region with \textit{Gaia} DR2 and APOGEE-2 \citep{Kounkel2018}, finding that the measured strength of turbulence for individual groups depends in part on their location relative to supernova epicenters \citep[see also][for similar results for the Perseus, Taurus, and Ophiuchus star-forming regions]{Ha2022}.

\subsubsection{\textbf{Vela Shell and the Vela OB2 Association}}
As illustrated for the Local Bubble and the Orion Big-Blast events, robust characterization of triggered star formation requires simultaneous constraints on the 3D positions, 3D space motions, and ages of its constituent stars. \citet{Bouy2015} computed the spatial density of OB stars using the Hipparcos catalog and recognized the Vela OB2 complex as one of the major structures of the solar neighborhood; however, the scarcity of precise proper motions, parallaxes, and radial velocities at that moment hampered a more detailed investigation of the internal structure of the complex itself. \citet{Jeffries2014} found two kinematically distinct populations A and B, around $\gamma^2$ Vel, using radial velocity measurements from the \textit{Gaia}-ESO survey \citep{Gilmore2012}. The subsequent study conducted by \citet{Sacco2015} detected additional stars with similar kinematic properties to the more dispersed population B previously identified by \citet{Jeffries2014}, supporting the existence of a young and dispersed population in the Vela OB2 complex.

Recent studies based on \textit{Gaia} put the complex structure under a new perspective. For example, \citet{Franciosini2018} showed that the two populations of the $\gamma$ Vel cluster are not only kinematically distinct but are also located at different distances. \citet{Beccari2018} identified a sextet of clusters in the Vela OB2 region based on the DBSCAN clustering algorithm \citet{Ester1996}, recovering the two previously known clusters from pre-\textit{Gaia} studies ($\gamma$ Vel and NGC 2547), and discovering another four sub-populations. Based on the color-magnitude diagram of these clusters, the authors concluded that four of them formed coevally 10 Myr ago while the other two clusters (including NGC 2547) have an age of 30 Myr. A traceback analysis of stellar trajectories based on the radial velocity information that is currently available for each cluster reveals no common location from where the clusters could have formed in the past.

However, the study conducted by \citet{Cantat-Gaudin2019b} increased the number of clusters in the Vela OB2 complex to eleven. The spatial distribution of the young stellar clusters forms a ring-like structure that appears to trace the edge of the IRAS Vela Shell. \citet{Armstrong2020} found firm evidence for anisotropic expansion in the complex by combining \textit{Gaia} DR2 astrometry with radial velocities from the \textit{Gaia}-ESO Survey. The observed spatial distribution of the young stellar clusters, their dynamics, and their age difference are likely to result from a triggered star formation episode that imprinted the expanding motion on the stellar population and the IRAS Vela Shell itself\index{Triggered star formation}. \citet{Cantat-Gaudin2019b} proposed a scenario in which stellar feedback and supernovae explosion in the 30 Myr old clusters swept the gas out of the central cavity and powered the expansion of the IRAS Vela Shell. Then, the second burst of star formation was triggered by the expansion of the IRAS Vela Shell $\approx$ 10 Myr ago and formed the younger stars in the cavity's rim. Of course, this scenario relies on the age difference and location of the various subgroups in the complex, and it connects the history of the stellar association with the IRAS Vela Shell. 

\subsubsection{\textbf{The Perseus-Taurus Supershell}}

No feedback-driven bubbles exist in isolation --- as more cavities are characterized in 3D, it will be possible to begin to constrain how the interaction between multiple intersecting shells regulates star formation across the solar neighborhood. One such interaction between superbubbles is shown in the right-hand panel of Figure \ref{fig:local_bubble}, which includes a small green sphere poking into the Local Bubble ``from the left” (the Galactic anti-center direction). That green sphere, known as the Perseus-Taurus Supershell, is another supernova-driven bubble discovered using \textit{Gaia}-based 3D dust mapping ---in this case, the \citet{Leike2020} map.
 
Perseus and Taurus are two nearby clouds whose 3D density structure was mapped out by \citet{Zucker2021}. When all the dust structures associated with Perseus and Taurus are viewed in 3D, it becomes clear that they curve around the surface of a nearly empty void about 150 pc across \citep{Bialy2021, Doi2021}. 

For decades, it was unclear whether the Perseus and Taurus clouds, which touch in projection on the plane of the sky, are connected in 3D \citep[see][]{Bialy2021, Doi2021}. The range of gas radial velocities present in Perseus and Taurus also overlap, but their average gas velocities (as well as the 3D space motions of their young stars) do not fit neatly into a picture of expansion \citep[see also][]{Kounkel2022}. 

However, if one looks closely at Figure \ref{fig:local_bubble}, especially the interactive versions, the Taurus cloud (marked red) is being squeezed by both the Local Bubble and the Perseus-Taurus Supershell. The influence of the Local Bubble's shell --- likely powered by more (and more recent) supernovae activity --- appears to be greater, as Taurus is moving today, away from the Sun and the center of the Local Bubble. Thus, signatures of expansion can be complicated by interacting bubbles, underlining the need for a more complete census of bubbles across the solar neighborhood, as well as constraints on their probable range of ages and the degree of energy injection necessary to account for their present-day morphologies. As these bubbles age, Galactic shear will also stretch and deform these bubbles, so a detailed accounting of both feedback \textit{and} Galactic shear will need to be taken into account in future work \citep[see e.g.][]{Li2022_bubbleshear}.

\begin{figure*}[h!]
\begin{center}
 \includegraphics[width=14cm]{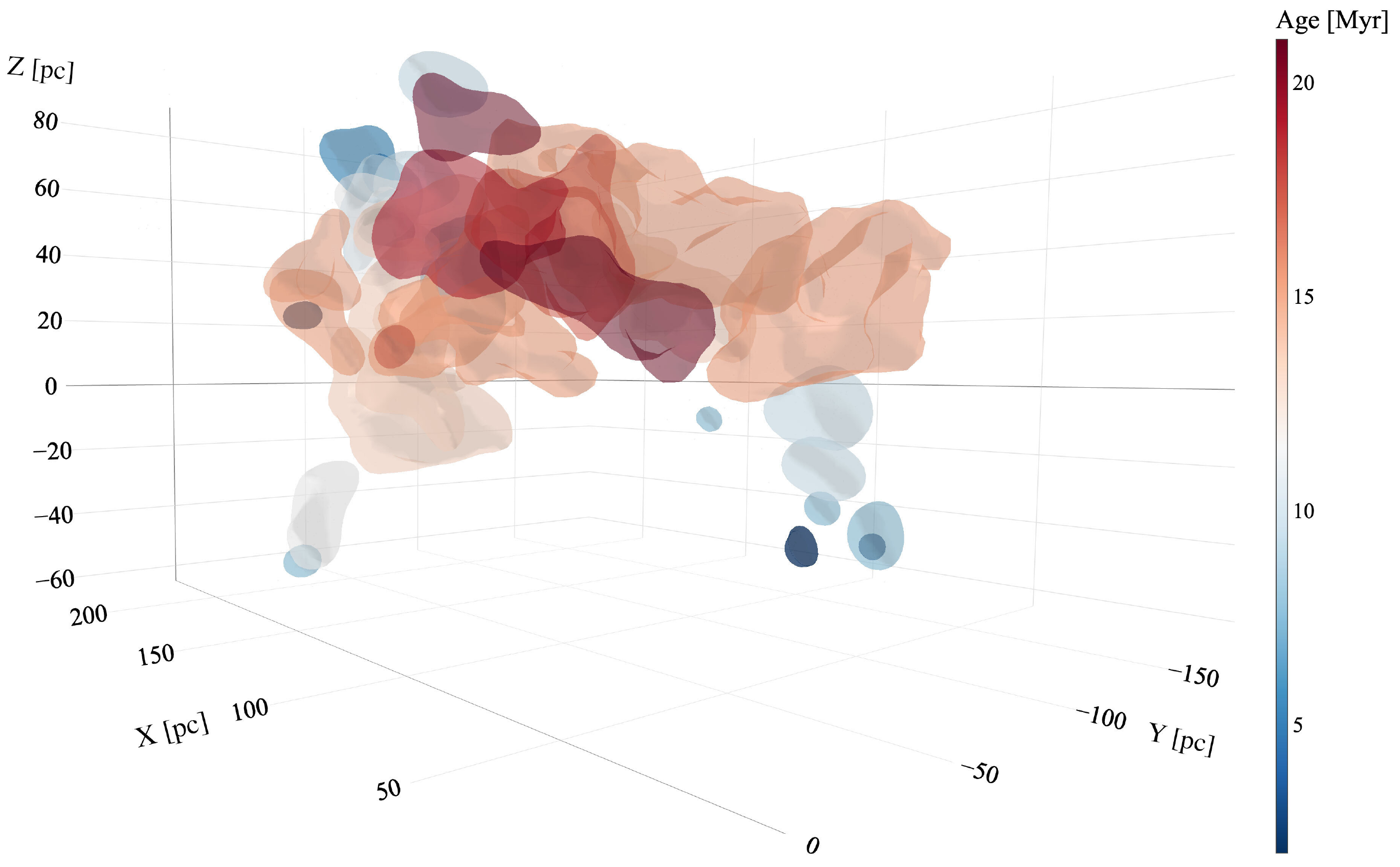}
 \caption{\small Cartesian 3D distribution of the newly identified stellar population in Sco-Cen. There are 34 coeval and co-moving clusters inside the Sco-Cen association. The color of the surfaces containing the different clusters encodes age, from dark blue (2 Myr) to dark red (21 Myr). All the star-forming regions in the vicinity of Sco-Cen, namely, Ophiuchus, L134/L183, Pipe Nebula, Corona Australis, Lupus, and Chamaeleon are part of Sco-Cen and are included in this figure. The central part of the association (UCL) is the oldest. Several systematic age gradients can be seen. Figure from \citet{Ratzenbock2023} based on the methodology of \citet{Ratzenbock2022-mr}. An interactive version of this figure is available \href{https://faun.rc.fas.harvard.edu/czucker/Paper\_Figures/PPVII/Ratzenbock2023.html}{\textbf{\fbox{here}}}.}
 \label{fig:scocen-age}
 \end{center}
\end{figure*}

\subsection{The closest young stars to the Sun}

For any imaging instrument, the closest young stars to the Sun offer the possibility for the highest resolution studies, making these nearby young stars prime targets for star and planet formation research. The closest embedded young stars to Earth (ages $\sim$ 1 Myr) are in the Ophiuchus and Taurus clouds (at a distance of about 140 pc), followed by the embedded stars in Corona Australis and Lupus (about 150 pc). For stars younger than 20 million years, the largest group within about 200 pc from Earth is the Sco-Cen OB association \citep{Kapteyn1914-tt}, which covers a substantial part of the Southern sky (about 80$\times$80 deg$^2$), running across the fourth quadrant of the Galaxy. Closer than Sco-Cen, several Young Local Associations (or young moving groups) have been identified \citep{Riedel2017-sb,Gagne2018a, Gagne2018b, Kerr2022_movinggroup}, the closest being the $\beta$-Pic Moving Group \citep{Barrado1999,Zuckerman2001-qk,MiretRoig2020, Crundall2019} at a mean distance of about 30 pc and an age of about 20 Myr.\index{young stars}

Unsurprisingly, Sco-Cen has been the target of many studies using \textit{Gaia} data. This association contains the entire stellar mass spectrum from O-stars to brown dwarfs, including regions of ongoing star formation such as Ophiuchus and Lupus, making it a prime laboratory for planet formation and early stellar evolution. Summarizing the Sco-Cen \textit{Gaia} results since DR2, \cite{Damiani2019} made a manual selection of over-densities in position, velocity, and HRD location to compile 11,000 young sources. \cite{Schmitt2021-jv} used eROSITA in combination with \textit{Gaia} to search for low-mass Sco-Cen members, finding 6,190 X-ray sources. \cite{Zerjal2021-wh} made use of a Bayesian tool to decompose stellar groups using the full 6D kinematic data, also performing an age determination. They identify eight distinct kinematic components containing in total 9,556 sources. \cite{Squicciarini2021-xd} studied 2,745 potential members in the Upper-Scorpius region by selecting subgroups solely based on kinematics. They divided the Upper-Scorpius region into eight groups: the clustered population (1,442 stars) and an older diffuse population (1,303). \cite{Miret-Roig2022} applied a Gaussian mixture clustering technique on \textit{Gaia} EDR3 for the Upper-Scorpius region. They include radial velocities, when available, to identify seven stellar groups containing 2,810 sources. \cite{Luhman2022-zc} investigated the region using a manual selection approach to \textit{Gaia} EDR3 data and identified 10,509 kinematic candidate members. \cite{Briceno-Morales2022-cr} combined the Convergent Point method with clustering techniques, complemented with age estimations, to find that Sco-Cen can be divided into three main kinematic structures, also finding seven stellar groups in Upper-Scorpius.\index{Clusters, structure}

Recently, \cite{Ratzenbock2022-mr} developed a new clustering method to extract co-spatial and co-moving stellar populations from large-scale surveys such as \textit{Gaia} using positions and proper motions (5D). They applied it to \textit{Gaia} DR3 data of Sco-Cen and found more than 13,000 co-moving young objects, a fifth of these having sub-stellar mass. Claiming higher sensitivity to small velocity differences, these authors recover 34 co-moving and coeval clusters in Sco-Cen (see Figure~\ref{fig:scocen-age}). The 3D distribution of these 34 coeval clusters implies a larger extent and volume for the Sco-Cen OB association than typically assumed in the literature. It confirms that the star-forming molecular clouds in the Sco-Cen region, namely, Ophiuchus, L134/L183, Pipe Nebula, Corona Australis, Lupus, and Chamaeleon, are part of the large Sco-Cen star formation event, which is also likely related to the formation of the Local Bubble \citep{Zucker2022} (see \S \ref{triggeredsf}).\index{young stars, ages}

Searching for young stars within 333 pc from the Sun, \citet{Kerr2021} identified in the \textit{Gaia} data 27 young groups. Ten have visible substructure, including notable young associations such as Orion, Perseus, Taurus, and Sco-Cen. They also identify earlier bursts of star formation in Perseus and Taurus that predate current, kinematically identical active star-forming events, suggesting that the mechanisms that collect gas can spark multiple generations of star formation, punctuated by gas dispersal and cloud regrowth. The distribution of the groups follows, remarkable, the location of the Radcliffe Wave and the Split, two Galactic-scale features seen in 3D dust, which will be discussed further in \S\ref{section5}.

Recently, \cite{McBride2021-cm} presented a trained neural network model that relies on \textit{Gaia} DR2 and 2MASS photometry to identify pre-main sequence stars and derive their age. Remarkably, they identified what appears to be a new feature of the local distribution of young stars: two concentric rings of young stars centered on the Sun, one at $\sim$100 pc with ages of up to $\sim $40 Myr, tracing the outer edges of the Local Bubble, and a second one at $\sim$300 pc with ages of $\sim $20 Myr. These authors associate the origin of the $\sim $40 Myr ring with the Local Bubble. These rings, if confirmed, would be remarkable because they are near-perfect circles of young stars centered on the Sun. One challenge to the double-ring interpretation is how 20-Myr and 40-Myr old rings with radii on the order of a few hundred parsecs could remain immune to the differential rotation of the Galaxy, which would naturally transform these rings into ellipses. Also, the 40-Myr ring does not match the age of the Local Bubble (about 14 Myr, see \S\ref{LocalBubble}) by at least a factor of two, so the Upper Centaurus Lupus and Lupus Centaurus Crux populations \citep[most likely responsible for the formation of the Local Bubble;][]{Breitschwerdt2016,Zucker2022} could not have caused the formation of the 40-Myr old ring. The fact that the larger of the two concentric rings is the youngest also complicates formation scenarios. Further exploration of the dynamics of these rings, potentially with \textit{Gaia} DR3 radial velocities, should not only provide insight into the plausibility of these formation scenarios, but should also settle the larger question of whether these rings are true physical structures. 

Ultimately, there is still much to gain from applying advanced machine learning tools to \textit{Gaia} data and extracting young stellar populations, but also much to learn about their limitations and the artifacts these tools might create. Still, particularly in the context of the next \textit{Gaia} Data Releases, constructing a high-spatial- resolution age map for the local Milky Way, as in Figure~\ref{fig:scocen-age}, is within reach.

\subsection{\textbf{Stellar streams and cluster coronae as the link between embedded systems and the Galactic field}}

Traditionally, the identification of physically connected stellar aggregates relied on locating spatial (2D) over-densities in the galactic field population \citep{Kapteyn1914-tt}. The Hipparcos catalog \citep{Perryman1997} offered the first possibility of kinematic profiling of stellar systems with a profound impact on our knowledge of co-moving groups of stars \citep{Piskunov2006}. The \textit{Gaia} mission elevated these prospects to a new level by enabling the identification of co-moving systems in velocity space, thereby mitigating the principal challenge of separating genuine members of an association or cluster from the often overwhelmingly abundant unrelated field stars.\index{Stellar groups}

\begin{figure*}[h!]
 \includegraphics[width=17cm]{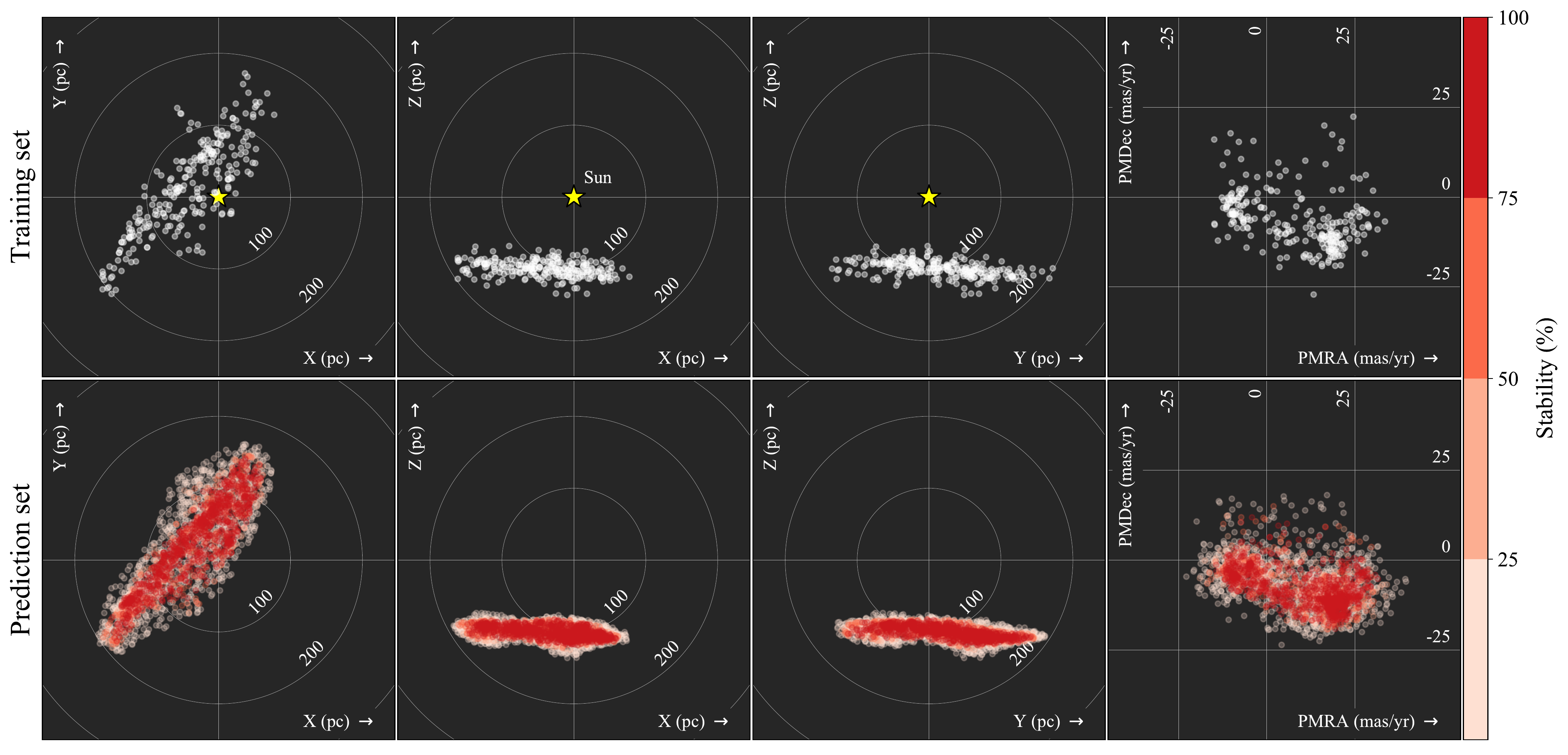}
 \caption{\small The Meingast-1 stream (sometimes referred to as the Pisces-Eridanus stream) is a Pleiades-age-and-mass population distributed across over 400 pc. Top row: Galactic Cartesian projection of the discovery data \citep{Meingast2019}, used as the training set for a clustering technique developed in \citet{Ratzenbock2020}. The yellow star represents the Sun. Bottom row: the recovered population contains about an order of magnitude more sources.
 }
 \label{fig:meingast1}
\end{figure*}

Using mainly kinematic data, \citet{Meingast2019} were able to identify a massive new type of stellar aggregate in the immediate vicinity of the Sun. The authors discovered a co-moving population of stars that --- despite being located at a distance of 100 pc and having a total mass greater than the Pleiades star cluster --- eluded discovery due to its sparseness in spatial density. Figure~\ref{fig:meingast1} displays the member selection of the system called Meingast-1 (sometimes referred to as the Pisces-Eridanus stream) by \citet{Ratzenbock2020} and reveals its nature as a several hundred parsec-long elongated streams of stars \citep{Roser2020}. Building on the original (kinematic) member selection, the authors used a sophisticated machine learning algorithm based on support vectors to compile a reliable membership catalog of the stream. The rightmost panels in this figure reveal the challenge of finding similar sparse stellar systems in proper motion space. In proper motions, the stream covers a large extent of the parameter space that contains hundreds of thousands of unrelated field stars.

Since the discovery of the stream-like structure, follow-up studies have measured rotation rates of main sequence stars with TESS data \citep{Ricker2014} and determined its age to be similar to the Pleiades, around 120 Myr \citep{Curtis2019}. The co-eval nature of the system was further confirmed by analyzing the correlation between lithium and stellar rotation rates \citep{Arancibia-Silva2020}. As such, the coeval nature of the stream and the Pleiades provide a uniquely valuable laboratory for testing various astrophysical processes under vastly different environmental conditions (clustered vs. non-clustered). As a prime example, the stream already serves as a testing ground for planet formation theory. A new planetary system has already been associated with one of its member stars \citep{Newton2021}.

\begin{figure*}[h]
 \includegraphics[width=17cm]{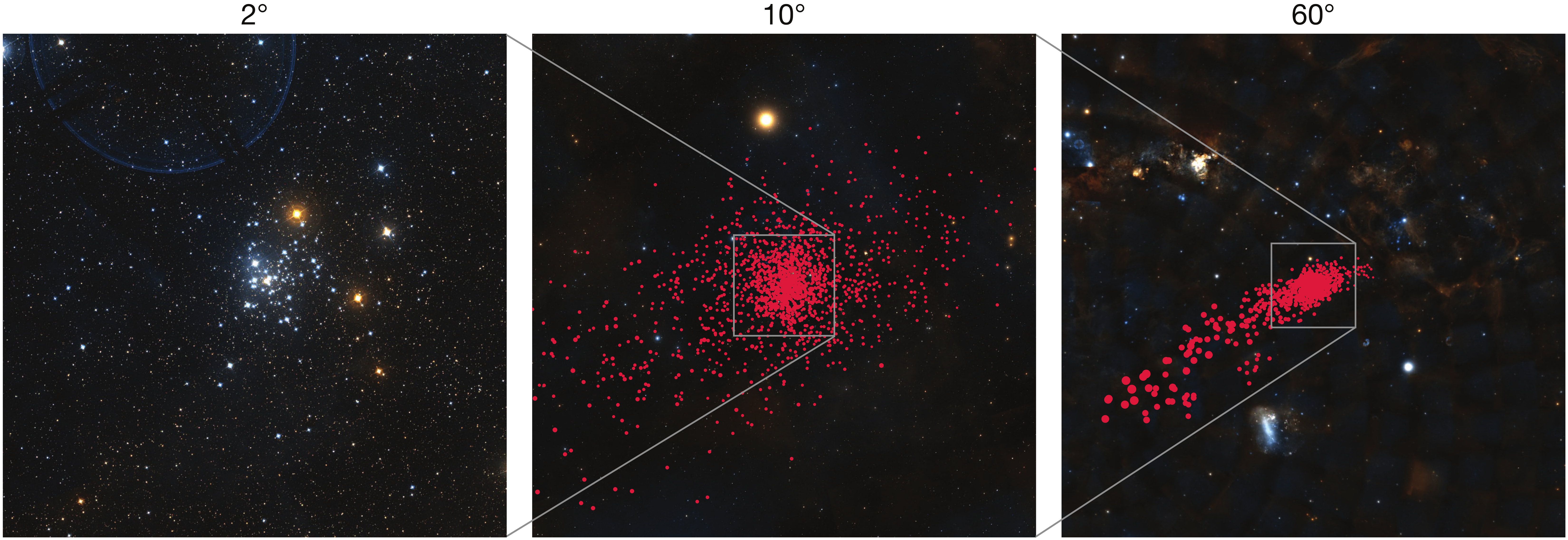}
 \caption{\small Cluster NGC 2516: the core (left panel) and the massive cluster corona (red dots, middle and right panels) as selected by \citet{Meingast2021}.}
 \label{fig:cluster_corona}
\end{figure*}

Other examples of similar young stellar structures conclude that such highly elongated systems could be relatively common in the solar neighborhood. \citet{Beccari2019} investigated young stars toward the Vela OB2 region and found stellar bridges across a few hundred parsecs that connect several known clusters in the area. The authors argue that their finding resembles the products of a star formation episode in a filamentary GMC about 35 Myr ago. Similarly, \citet{Jerabkova2019} identify a 17 Myr old and $\approx$ 90 pc long stellar relic filament in the Orion star-forming region but argue that the structure is too young to have been entirely shaped by tidal forces, but rather is also a remnant of a filament turned to stars. \citet{Tian2020} presented another example of such highly elongated structures by identifying a massive 40 Myr old, 200 pc long stellar system in the direction of the Orion region, however, with kinematics connecting it rather to the Vela OB2 complex \citep[see also][]{Wang2022_snake}. The young stellar filaments in \citet{Beccari2019,Jerabkova2019,Tian2020} may be related to the hundreds of predominantly older stellar ``strings" presented in \citet{Kounkel2019,Kounkel2020_strings}, who also argue that the filamentary morphology of these structures is primordial \citep[see also][for analysis of the chemical homogeneity of the strings]{Manea2022}. However, follow-up work by \citet{Zucker2022_strings} --- examining the radial velocity dispersions of their stellar members with updated \textit{Gaia} DR3 data --- finds that many of these gravitationally unbound strings have large velocity dispersions, averaging $\rm \approx 16 \; km\; s^{-1}$. In contrast to streams like Meingast-1 (with velocity dispersion of $\rm 1.3 \; km \; s^{-1}$), the large velocity dispersions of the strings imply that their reported ages are orders of magnitude larger than their predicted dispersal times, calling into question the physicality of a large fraction of the sample and underlining the need for dedicated follow-up studies on individual systems \citep[see e.g.][]{Andrews2022}. All the above-listed findings highlight the complexity of young stellar systems in the solar neighborhood and further show our incomplete knowledge regarding the fidelity and formation of such filamentary stellar structures. 

Even beyond structures like stellar relic filaments, snakes, and strings --- all systems claimed to be co-eval --- \textit{Gaia} has also enabled the discovery of stellar ``pearls", or distinct clusters that follow similar orbits in the Galaxy and manifest as overdensities in action-angle space \citep{Coronado22}. While these pearls are not claimed to be born at the same time within the same parental molecular gas structure, the existence of pearls suggest that recent star formation in the disk is strongly clustered towards a small set of orbits, a finding that can no doubt be explored further given the deluge of new radial velocity data from \textit{Gaia} DR3. 

Naturally, the known star cluster population in the solar vicinity has also been subject to several studies. Many of these focused on identifying new clusters and investigating existing ones \citep{Cantat-Gaudin2018, Cantat-Gaudin2020, Castro-Ginard2020, Castro-Ginard2022, Dias2021, Hao2022, He2022a, He2022b, Hunt2021, Liu2019, Sim2019, Jaehnig2021, Fu2022}. The general conclusion of these efforts was that some previously found clusters were now shown to be likely physically non-existent while a whole trough of new clusters was cataloged \citep[e.g.,][]{Kos2018}. \textit{Gaia} also enabled the discovery of tidal features that protrude from some of the older clusters in the solar neighborhood \citep{Bhattacharya2022, Boffin22, Casamiquela22, Furnkranz2019, Li2021, Jerabkova21, Meingast_Alves2019, Pang2021, Roser2019a, Roser2019b, Gagne2021}.

The complexity of young stellar systems in the solar neighborhood was especially highlighted by analyzing the structure of nearby young clusters\index{OB associations}. Specifically, clusters associated with OB associations appear as large compounds and sometimes show complex networks of young stars \citep{Armstrong22, Beccari2019, Cantat-Gaudin2019a, Chemel22, Kounkel2018, Miret-Roig2022, Orellana2021, Quintana21, Squicciarini2021-xd, Pang2022, Ratzenbock2022-mr}. \citet{Meingast2021} reanalyzed a set of ten nearby young clusters\index{Stellar clusters}. The authors showed that so-called stellar coronae envelope all clusters in their sample --- several-hundred-parsec-long halos surrounding the otherwise easily discernible cluster core. They also showed that, for their sample of clusters, most of the stellar mass of these systems is not gravitationally bound but is rather distributed in a cluster’s corona. Their analysis further revealed that these systems show signs of a tidal influence on the surrounding galactic field. An example of the vast extent of these coronae is shown in Figure \ref{fig:cluster_corona}, which shows a view of the Milky Way centered on cluster NGC 2516 from different perspectives. The left panel shows a field of $2^\circ$ with the cluster core readily discernible from the galactic field. The panels in the center and on the right are also centered on the cluster but show a field of view of $10^\circ$ and $60^\circ$, respectively. In addition, these panels also show the newfound members of NGC 2516 and distinctly highlight the immense size of the population across the sky. In a follow-up study, \citet{Bouma2021} leveraged a combination of TESS, \textit{Gaia}, GALAH, and \textit{Gaia}-ESO data to show that the isochronal, rotational, and lithium age of stars in NGC 2516's halo are consistent with those in its core, validating the existence of the stellar corona. 

From a theoretical point of view, a significant number of stars can escape the gravitational potential of a cluster during its transition from an embedded state to a gas-free state \citep{Baumgardt2007, Dale2015}. The expansion depends on several factors, such as the star formation efficiency \citep{Lada1984} and the timescales of gas expulsion \citep[e.g.,][]{Dinnbier2020a, Dinnbier2020b, Pang2020}. At the same time, the shape of the stellar coronae observed today could be a consequence of the complex morphology of star-forming regions. Young star clusters are found to be embedded in a complex network of often actively star-forming filamentary structures\index{Stellar clusters}. The sample of clusters analyzed in \citet{Meingast2021} does not show a correlation between the coronae's size and the host cluster's age. This suggests that both stars evaporating from the cluster, and stars that formed in the surrounding environment together, comprise the population of the stellar coronae.

\section{\textbf{THE SOLAR NEIGHBORHOOD IN A GALACTIC CONTEXT}} \label{section5}

\begin{figure*}[ht!]
 \includegraphics[width=17cm]{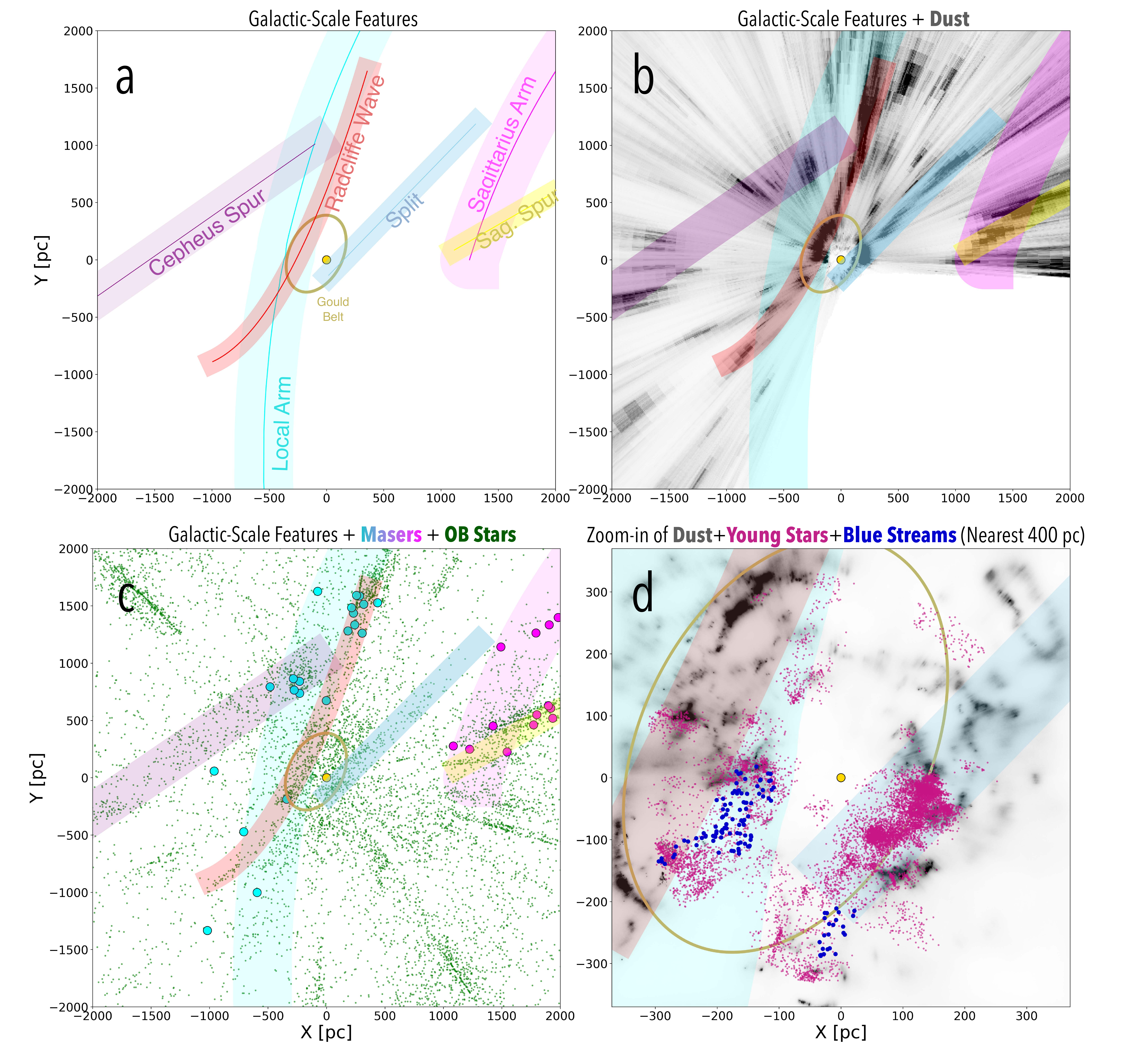}
 \caption{\small Galactic structure in the solar neighborhood. (a): Bird's-eye view of the nearest 2 kpc, with colored stripes showing large-scale features which are labeled by name. Features include the Cepheus Spur \citep{Pantaleoni_Gonzalez2021}, the Radcliffe Wave \citep{Alves2020}, the Local Arm \citep[as defined in][]{Reid2019}, the Sagittarius Arm \citep[as defined in][]{Reid2019}, the Split \citep{Lallement2019}, and the Sagittarius Spur \citep{Kuhn2021b}. A model for the Gould Belt, whose existence has faced substantial challenges in the \textit{Gaia} era, is also overlaid as a brown ring \citep{Perrot2003}. (b): Large-scale features overlaid on the 3D dust map from \citet{Green2019}. (c): Large-scale features over-plotted with masers corresponding to the Local (cyan points) and Sagittarius (magenta points) arms \citep{Reid2019}, alongside the distribution of OB stars from \citet{Pantaleoni_Gonzalez2021} (green points). (d): Zoom-in of the nearest 400 pc, showing the 3D dust from \citet{Leike2020}, the Blue Streams of OB stars \citep{Bouy2015} (blue points), and the distribution of young stellar clusters $\rm \lesssim 50 \; Myr$ old (deep violet-red points) from \citet{Kerr2021}. Note that the distribution of young stars from \citet{Kerr2021} broadly follows two paths: the location of the Radcliffe Wave (red band) and the Split (blue band), two large-scale 3D dust features. An interactive version of the datasets shown in panels (a) - (c) is available \href{https://faun.rc.fas.harvard.edu/czucker/Paper_Figures/PPVII/Synthesis.html}{\textbf{\fbox{here}}}.}
 \label{fig:synthesis}
\end{figure*}

One of \textit{Gaia}'s most important legacies is the context it brings to our understanding of fundamental astrophysical processes in the near Galaxy. Understanding the 3D spatial distribution and connections between gas and young stellar structures, as well as their kinematics, matters when disentangling complex star and planet formation regions. This 3D context is critical to directly characterize stellar feedback and high-energy processes, as well as measure the roles of gravity, magnetic fields, turbulence, and galactic dynamics in structure formation. This section focuses on the distribution of gas and young stars in a galactic context. We start by discussing nuances and revisions to canonical spiral structure theory as revealed by \textit{Gaia} mapping of Upper Main Sequence (UMS) stars, OB stars, and open clusters. We then review evidence for --- and significant opposition against --- a long-standing structure in the solar neighborhood known as Gould's Belt. We proceed to reviewing some of the unexpected large-scale features recently discovered in the \textit{Gaia} era, which form connections between star-forming regions previously thought to be isolated. A roadmap of these new Galactic-scale features discovered in the \textit{Gaia} era is summarized in Figure \ref{fig:synthesis}. We conclude by discussing how this emerging \textit{Gaia}-based model of our solar neighborhood facilitates new comparisons with both numerical simulations and extragalactic observations of Milky Way-like galaxies. 

\subsection{\textbf{\textit{Gaia} and Spiral Structure}} \label{spiral}

The first hint of spiral structure near the Sun was found in the seminal work of Morgan and collaborators \citep{Morgan1953} using the location of HII regions to infer the distribution of short-lived high-mass stars and the location of three spiral arms\index{HII regions}. These three spiral arms, namely, the Sagittarius arm, the Orion (or Local) arm, and the Perseus arm, have been extensively characterized in the literature using the latest results from maser surveys \citep{Reid2019}, two of which (Sagittarius and Orion/Local) have significant segments lying within 2 kpc of the Sun. The Local and Sagittarius arms as defined by \citet{Reid2019} are shown via the cyan and magenta traces in Figure \ref{fig:synthesis}.

Recently, \citet{Poggio2021}, using \textit{Gaia} data, mapped the density variations in the distribution of young upper main sequence (UMS) stars, open clusters, and classical Cepheids in the Galactic disc within several kiloparsecs of the Sun. The resulting overdensity maps exhibit large-scale arches that extend in a clumpy but coherent way over the entire sampled volume, showing the location of potential spiral arm segments near the Sun. Peaks in the UMS overdensity are well-matched by the distribution of young and intrinsically bright open clusters. While the resulting map based on the UMS sample is generally consistent with previous models of the Sagittarius-Carina spiral arm, the geometry of the arms in the III quadrant (Galactic Longitudes $180^\circ < l < 270^\circ$) differs significantly from that suggested by many previous models. In particular, their map favors a larger pitch angle for the Perseus arm, and the Local Arm extends into the III Quadrant at least 4 kpc past the position of the Sun, giving it a total length of at least 8 kpc.

At about the same time, \citet{Zari2021} constructed a sample of hot, luminous stars using \textit{Gaia} data to find that the most clear overdensities in the distribution of hot stars can be associated with the presumed spiral arms of the Milky Way, in particular, the Sagittarius-Carina and Scutum-Centaurus arms. Similar to \citet{Poggio2021}, \citet{Zari2021} suggests a distribution of the young stellar structures on the Milky Way's disk that does not seem to match well the canonical spiral arm structure. This is another \textit{Gaia} surprise that will have fundamental implications on establishing what type of spiral galaxy the Milky Way is --- grand design or flocculent, or likely, in between --- currently an open topic that goes beyond the scope of this review. 

While Poggio et al. and Zari et al. focused on massive stars as tracers of Galactic structure, an equally recent paper by \citet{Castro-Ginard2021} used overdensities of open clusters younger than 30 Myr in the Perseus, Orion (Local), Sagittarius, and Scutum spiral arms. These authors used the birthplaces of the open cluster population younger than 80 Myr to trace the evolution of the different spiral arms and compute their pattern speed. They could increase the range in Galactic azimuth by adding 264 young open clusters to the 84 high-mass star-forming regions used so far to find that spiral arms nearly co-rotate with field stars at any given radius, discarding a common spiral pattern speed for the studies’ spiral arms. They find that the different spiral pattern speeds disfavor classical density waves as the key drivers for forming the Milky Way spiral structure. Their findings seem to agree better with simulation-based approaches that favor transient spirals \citep[see also][]{Sellwood2019}. With four times more stellar radial velocities now available with \textit{Gaia} DR3 than utilized in the \citet{Castro-Ginard2021} study, the coming years should provide further clarity on the nature and persistence of spiral structure both in the solar neighborhood and across the Galaxy at large. 

\subsection{\textbf{Challenges to the Gould’s Belt}}

On intermediate scales --- independent of spiral arms --- our understanding of Galactic structure locally for the past 150 years has been shaped by a feature known as the “Gould's Belt" \citep{Gould1874, Palous2016} a flattened disk of interstellar material with a diameter of about a kiloparsec, inclined to the plane and containing essentially all nearby massive stars, plus a few million solar masses of gas \citep{Blaauw1952, Lindblad2000, Perrot2003}. A model for the Gould's Belt is overlaid in brown in Figure \ref{fig:synthesis}. Traditionally, the Gould’s Belt has contained all the nearby star-forming regions: Chamaeleon, Lupus, Pipe, Corona Australis, Ophiuchus, Cepheus, Aquila Rift, Serpens, and Perseus \citep{Dzib2018, Loinard2012}. Its assumed kinematics differ from the motions of older stars, and its velocity field showed motion that significantly deviated from rotation around the distant center of the Milky Way. Historically, the Gould’s Belt was perceived as expanding and revolving around a nearby center toward the Cassiopeia–Taurus OB association \citep{Blaauw1991-mx}. However, no convincing model exists to explain Gould’s Belt \citep{Palous2016}. Here, we summarize recent literature which uses \textit{Gaia} data to argue in favor of the Gould's Belt, including how analysis of the 3D motion of young stars towards Gould's Belt clouds provides kinematic evidence of the Belt's expansion. However, we also summarize significant opposition to the existence of the Gould's Belt, as revealed by unbiased surveys of stars and dust in the solar neighborhood. 

\citet{Dzib2018} compiled \textit{Gaia} data on young stellar objects for a subset of star-forming regions associated with the Gould’s Belt. Using the 3D spatial distribution of these YSOs, Dzib et al. determined the shape of the Gould’s Belt, parameterizing it as an ellipsoid with a size of $\approx$ 350 pc in $x$ and $y$ and $\approx$ 70 pc in $z$, centered about 100 pc from the Sun. Combining the 3D spatial positions of the YSOs with \textit{Gaia} proper motions and ancillary radial velocity measurements from the literature, Dzib et al. argue that the Gould Belt is expanding at a rate of $\rm 2.5 \; km \; s^{-1}$. \cite{Bobylev2020-ez} analyzed the spatial and kinematic properties of a large sample of young T Tauri type stars in a 500 pc radius from the Sun and derived updated physical parameters for the Gould's Belt. Using a sample of 100,000 red clump giants, \citet{Gontcharov2020} analyzed the 3D spatial dust distribution in the Gould’s Belt, finding it is also well-described by an ellipse but about double the size as constrained by the YSOs in Dzib et al. \citet{Palous2019} perform a study using a \textit{Gaia}-detected sample of OB stars, finding that OB associations are expanding locally, but with no evidence of a common origin.

While several studies in the \textit{Gaia} era find evidence of the Gould's Belt in both gas and young stars, most of these are targeted studies focused on regions commonly associated with the Belt. However, \textit{Gaia} has also enabled dozens of new, unbiased maps that adopt a more broad-brush approach to Galactic structure, targeting all gas and stars within the nearest few kiloparsecs. None of these maps support the existence of the Gould's Belt. In fact, already during the Hipparcos era, pre-\textit{Gaia}, both \cite{Elias2009-se} and \cite{Bouy2015} raised doubts on the existence of the Gould's Belt. Using \textit{Gaia} data, \citet{Zari2018} construct a uniform three-dimensional density map of young stars out to 500 pc. Zari et al. ``find that the 3D density maps show no evidence for the existence of the ring-like structure, which is usually referred to as the Gould Belt." There is no ``link" between opposite sides of the Belt, either in gas or young stars. The lack of evidence for the Gould's Belt in unbiased studies of young stars is also supported by the catalog of \citet{Prisinzano2022}, who use DBSCAN in combination with \textit{Gaia} EDR3 parallaxes, proper motions, and broadband photometry to identify a uniform sample of young stars $\rm < 10 \; Myr$ old within 1.5 kpc of the Sun.

\subsection{\textbf{Unexpected Large-Scale Features in the \textit{Gaia} Era}} 

Rather than affirming traditional models of the local solar neighborhood, new 3D dust maps, new catalogs of molecular cloud distances, and new maps of the 3D spatial distribution of OB stars and young stellar clusters provide evidence for structures much larger than Gould's Belt and challenge existing models for spiral structure at high resolution. These features, shown in Figure \ref{fig:synthesis}, include the 2.7 kpc “Radcliffe” Wave containing most of the Gould Belt clouds, a 2 kpc-long “Split” extending out of the Sco-Cen association, prominent spurs between the Sagittarius/Local arms and Perseus/Local arms defined by the distribution of OB stars, and a 1-kpc-long high-pitch-angle structure in the Sagittarius arm defined by the 3D spatial distribution of young stellar clusters.

\begin{figure*}[ht!]
 \includegraphics[width=17cm]{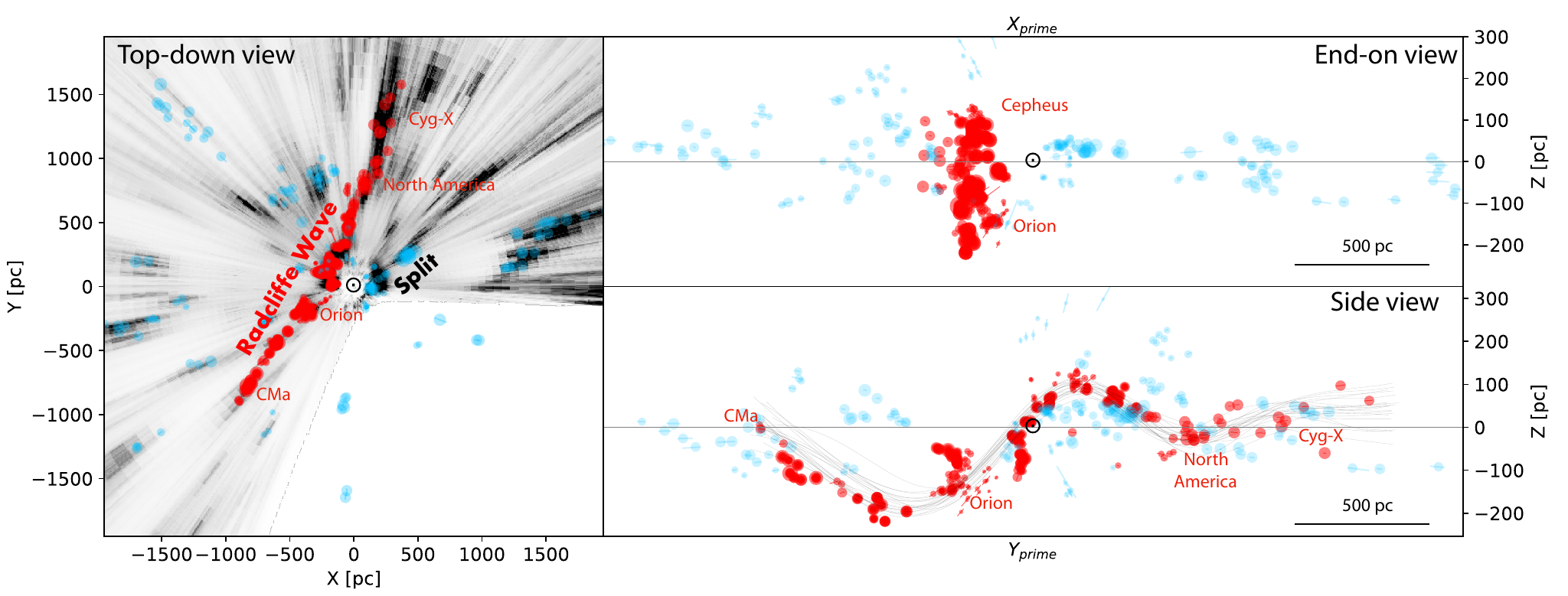}
 \caption{\small 3D distribution of local clouds. The position of the Sun is marked with $\odot$. The size of the symbols is proportional to the column density.
The red points describe a spatially and kinematically coherent structure, the Radcliffe Wave. Remarkable, and currently unexplained, the structure appears to be undulating on the Galactic plane. The greyscale map in the left panel show an integrated dust map. For an interactive version of this figure, including additional layers not shown here (for example, a model of the Gould Belt and log-spiral arm fits), see \href{https://faun.rc.fas.harvard.edu/czucker/Paper_Figures/radwave.html}{\textbf{\fbox{here}}}. Reproduced from \cite{Alves2020}, \textit{Nature}, 578, 237.}
 \label{fig:radwave}
\end{figure*}

\subsubsection{\textbf{The Radcliffe Wave and the Split}} \label{rw_split}
There is mounting evidence that the star-forming regions traditionally associated with Gould's Belt clouds are only a small part of two linear kpc-long gas structures: the Radcliffe Wave \citep{Alves2020} and the Split \citep{Lallement2019}. The Sun is currently located in between these kpc-long structures. In Figure \ref{fig:synthesis}, we show the Radcliffe Wave in red and the Split in blue. The Radcliffe Wave and the Split were initially detected with 3D dust mapping, using either the global distribution of molecular clouds with accurate dust-based distances (Figure \ref{fig:cloudcatalogs}) or in gross topological maps of the interstellar medium (Figure \ref{fig:dustmap}).

The larger of the two local structures, the Radcliffe Wave, has an aspect ratio of about 1:20, extends for about 3 kpc and contains about three million solar masses of gas (see Figure \ref{fig:radwave}). Inside it, star-forming regions are connected by lower-density gas. The Radcliffe Wave contains four of the major Gould’s Belt clouds (Perseus, Taurus, Orion, and Cepheus) as well as several other famous star-forming regions, including CMa OB1, Mon R2, Cygnus X, and the North America Nebula. The Radcliffe Wave appears to be undulating and is well described by a damped sinusoidal wave on the plane of the Milky Way with an average period of about 2 kiloparsecs and a maximum amplitude of about 160 parsecs. The Radcliffe Wave is likely the densest gas distribution of the Local Arm \citep{Reid2019} (cyan trace in Figure \ref{fig:synthesis}), containing the bulk of its gas in the nearest 2 kpc, placing the Sun at about 100 pc from the inner side of a major arm of the Milky Way. However, it has a higher pitch angle ($\approx 30^\circ$) than traditionally associated with the Local Arm, estimated to be $11^\circ$ in \citet{Reid2019}. 

\begin{figure*}[ht!]
\centering
 \includegraphics[width=18cm]{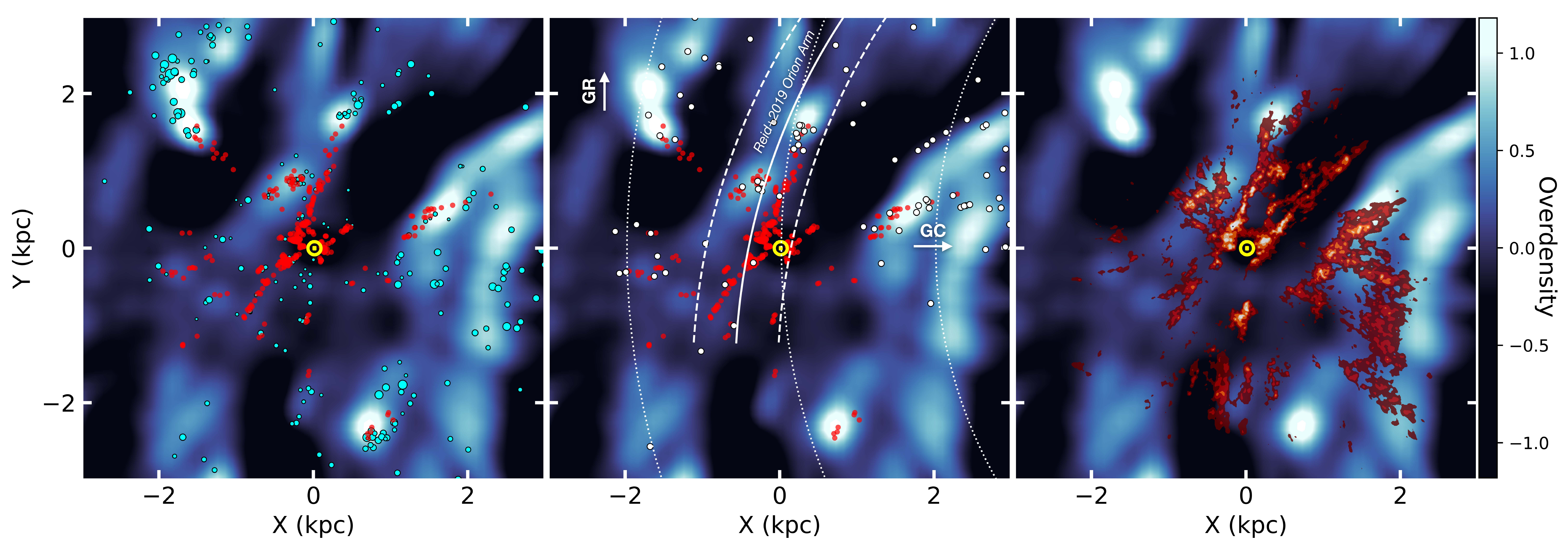}
 \caption{\small A Galactic XY (bird’s-eye) view of various stellar/dust distributions within a few kiloparsecs of the Sun (centered yellow dot). The over-density map of the OBA-type stars from \citet{Zari2021} (black-blue-white) is shown in the background of each panel. The emergent spiral features in this map, from decreasing to increasing values in X, are the Perseus, Orion, and Sagittarius-Carina arms. Star-forming clouds from \citet{Zucker2020} are displayed as red points in the first two panels, with the Radcliffe Wave appearing as a narrow alignment of clouds from ($x$, $y$) $\approx$ (-1, -1) kpc to (X, Y) $\approx$ (0.5, 2) kpc. The cyan points in the first panel are the open clusters from \citet{Cantat-Gaudin2020} with ages less than 30 Myr, sized relative to their membership count. The second panel shows the fit (solid white line) to the Orion arm from masers \citep{Reid2019}, three circles of constant, Galactocentric radius (white dashed lines), and two arrows pointing in the directions of the Galactic center/Galactic rotation. Overlaid in the right-most panel is the dust map of \citet{Lallement2019} with a heat map color-scale showing increasing dust density. Credit: Swiggum et al., A\&A, 664, L13, 2022, reproduced with permission \copyright ESO.
 }
 \label{fig:radwave_oba}
\end{figure*}

An obvious question now is, what is the relationship between the young populations used to define the spiral arms in \S \ref{spiral} \citep{Castro-Ginard2021, Poggio2021,Zari2021} and kpc-long gas structures like the Radcliffe Wave emerging from \textit{Gaia} 3D dust maps \citep{Alves2020, Green2019, Lallement2019, Zucker2020}? The answer, as we will see, is not obvious. In one study, \citet{Quillen2020} estimates the birth location of young $(\rm < 70 \; Myr)$ stellar clusters by performing backward orbit integrations in a gravitational potential, finding that the variation in the birth heights of these stellar associations suggest that they were born in a corrugated disk of molecular clouds, similar to those inferred from current 3D dust maps. In another study, \citet{Swiggum2022}, compared the distribution of massive stars from the literature \citep{Castro-Ginard2021, Pantaleoni_Gonzalez2021, Poggio2021, Zari2021} for the Orion (Local) arm, with the location of the Radcliffe Wave, and found that the dust and massive stars do not overlap exactly. In Figure \ref{fig:radwave_oba}, we present the distribution of young stars from Zari et al. where we overlay the Radcliffe Wave. In this figure, the location of the Radcliffe Wave is roughly parallel but shifted to the inner Galaxy compared to the massive stars and clusters in the Orion (Local) arm. This is similar to the morphology of spiral arms in nearby galaxies and suggests that the Radcliffe Wave could be the gas component of the Orion (Local) arm. If this is the case, the spatial distribution is tantalizingly close to the classical scenario for arm formation via a spiral shock \citep{Elmegreen2018}, but this result is in tension with \cite{Castro-Ginard2021}, which disfavors classical density waves as the drivers of spiral structure. In the spiral shock scenario, the enhancements seen in Poggio and Zari et al. would correspond to the slightly older populations (ages $<$ 50 Myr) in front of the currently star-forming gas of the Radcliffe Wave. In this scenario, one would expect an age (color) gradient across the arm, but this was not observed (nor ruled out). 

The origin of the undulation in the Radcliffe Wave is also unclear but could arise internally in the disk (e.g., because of the Kelvin-Helmholtz instability \citep{Fleck2020} or externally due to a collision \citep[e.g.][]{TepperGarcia2022}. To elucidate its origin, the Radcliffe Wave has been the focus of several studies seeking to characterize its kinematics. In the first of such studies, \cite{Donada2021-sc} presented a study of 13 clusters that are probable members of the Radcliffe Wave and found that the vertical motion of 11 of these ``is not contradictory with the behaviour expected from a simple model of harmonic motion in the vertical direction." \citet{Li2022} used the proper motions of YSOs associated with Radcliffe Wave to estimate that the Radcliffe Wave is ``possibly oscillating around the Galactic disk mid-plane with an amplitude of 130$\pm$20 pc''. \cite{Bobylev2022-tx} confirmed the existence of the Radcliffe Wave via the position of young stars. They also recognized the Radcliffe Wave in the positions and in the vertical velocities of masers and radio stars belonging to the Local Arm but argue that the Radcliffe Wave is less of a wave but ``more like a local high-amplitude burst, rapidly fading away." 

\citet{Thulasidharan2022} investigate whether the Radcliffe Wave may be correlated with a kinematic oscillation perpendicular to the Galactic disk. Leveraging a combination of open clusters, massive stars, and red giant stars in and beyond the Radcliffe Wave, \citet{Thulasidharan2022} find evidence of an oscillation in the vertical velocity of a few $\rm km \; s^{-1}$ in the younger populations, suggesting that the amplitude of oscillation is tied to the age of the stellar population. Running simulations of a Sgr-like dwarf galaxy impacting the Milky Way, \citet{Thulasidharan2022} find that the resulting density wave aligns with the Radcliffe Wave but that the kinematic wave is misaligned, underlining the need for further studies on the effects of a perturber on the Galactic disk. Following up on the study of \citet{Thulasidharan2022}, \citet{Tu2022} investigate the vertical angle of pre-main sequence stars in \textit{Gaia} DR2 along the wave --- corresponding to their orbital phase perpendicular to the Galactic plane --- finding that the vertical angle varies significantly from trough to peak, in a pattern potentially in agreement with a wavelike-oscillation. In an analysis of clustered star formation in the solar neighborhood, \citet{Alfaro2022} finds that warp, corrugations, and high local deviations in both $z$-height and vertical velocity all likely play an interconnected role in shaping the phase-space structure of young open clusters near the Sun, and better understanding the interplay of these phenomena will also likely give rise to better understanding of the formation of the Radcliffe Wave.\index{Clustered star formation} 

Towards the Galactic center, the remaining Gould Belt clouds in the solar vicinity (Ophiuchus, Lupus, Chamaeleon, Corona Australis, and the Pipe Nebula) belong to the newly discovered ``Split”, which \citet{Lallement2019} argue could be a spur-like feature between the Sagittarius and Local arms with a pitch angle of $\approx 45^\circ$ The Split begins in the famous Sco-Cen association and extends almost 2 kpc in length, with a similar width as the Radcliffe Wave. Its kinematics have yet to be systematically investigated in the \textit{Gaia} era, but morphologically it shares many properties with the newly-discovered Sagittarius spur \citep{Kuhn2021b}, as discussed in \S \ref{sagspur}.

While much work remains, what seems clear is that emerging kpc-scale linear features like the Radcliffe Wave and the Split detected in 3D dust maps are the ideal laboratory to test different spiral arm formation theories. The expected increase of astrometric precision in the subsequent \textit{Gaia} data releases and better radial velocities from ground-based surveys promises to settle this 70-year-old question regarding the nature of spiral structure.

\subsubsection{\textbf{The Blue Streams}}
\citet{Bouy2015} constructed a 3D map of the spatial density of OB stars within 500 pc from the Sun using the Hipparcos catalog and found three large-scale stream-like structures they called Blue Stream, for being composed of young stars (shown in blue in Figure \ref{fig:synthesis}d). The spatial coherence of these blue streams and the monotonic age sequence over hundreds of parsecs suggest they are made of young stars, similar to the young streams that are conspicuous in nearby spiral galaxies. The three streams are 1) the Scorpius to Canis Majoris stream (Sco-CMa), covering 350 pc and 65 Myr of star formation history; 2) the Vela stream, encompassing at least 150 pc and 25 Myr of star formation history; and 3) the Orion stream, including the well-known Orion OB1abcd associations. Their map reveals a remarkable and previously unknown nearby OB association, between the Orion stream and the Taurus molecular clouds (that they name Taurion), which they argued could be responsible for the observed structure and star formation activity in the Taurus complex. Post-\textit{Gaia}, it became clear that the Orion Blue Stream is part of the Radcliffe Wave. The Sco-CMa Blue Stream, minus the CMa region, which is part of the Radcliffe Wave, is likely associated with the Split. The Sco-CMa Blue Stream age gradient would then suggest an ordered transformation of the Split gas into stars, although more work is needed to test this hypothesis. 

\subsubsection{\textbf{The Sagittarius Spur}} \label{sagspur}

 The Sagittarius spur \citep{Kuhn2021b} is a 1-kpc-long+ band of star-forming regions, shown in yellow in Figure \ref{fig:synthesis}. The Spur was originally detected using the \textit{Spitzer}/IRAC Candidate YSO (SPICY) catalog \citep{Kuhn2021a}, which leveraged mid-infrared photometry to classify over 100,000 young stellar objects toward the inner Galactic plane. These YSOs were later grouped using a clustering algorithm on the plane-of-the-sky, before being cross-matched with \textit{Gaia} to get distance and proper motion information for each group. The Sagittarius spur contains 25 such YSO groups in the SPICY catalog, which together form a linear structure with a pitch angle of 56$^\circ$, significantly different from the traditional view of the Sagittarius spiral arm defined by masers, which \citet{Reid2019} argue has a pitch angle of 17$^\circ$. The Sagittarius Spur also contains about half of the masers used to define the traditional global log-spiral fit to the near Sagittarius arm in Reid et al (magenta trace in Figure \ref{fig:synthesis}) within the nearest 2 kpc. Like the Radcliffe Wave and the Split, the Sagittarius spur has a high aspect ratio, of $\approx$ 7:1 and links many of the most famous star-forming regions toward the inner Galaxy, including M8, M16, M17, and M20. 

While originally detected using YSO groups, the Sagittarius spur is also clearly seen in 3D dust, with the groups coinciding with a prominent dust lane showing the same high-pitch angle. While nominally 1 kpc in length (starting in the first quadrant and ending at the Galactic center) it is possible that the Sagittarius Spur extends even further into the fourth quadrant, comprising what \citet{Chen2020a} calls the “Lower Sagittarius-Carina Spur” as detected in their 3D molecular cloud catalog, shown in Figure \ref{fig:cloudcatalogs}.

Associating the proper motions of the spur's young stellar objects with its radial velocities obtained from CO emission of their natal clouds, \citet{Kuhn2021b} find the structure to be remarkably kinematically coherent, moving slightly faster than Galactic rotation and drifting marginally radially inward, with a bulk motion toward the negative $z$ direction. \citet{Kuhn2021b} calculate that the rotational shear experienced by the spur is, $\Delta \Omega$ is $\rm 4.6 \; km \; s^{-1} \; kpc^{-1}$, implying a relatively short predicted timescale of $\approx 90$ Myr for the structure to alter its pitch angle by a factor of two \citep{Elmegreen1980}.

\subsubsection{\textbf{The Cepheus Spur}}

Finally, an additional spur in the \textit{Gaia} era has been identified using new catalogs of OB stars. \citet{Pantaleoni_Gonzalez2021} cross-match the “Alma Luminous Star” catalog \citep{Cameron_Reed2003} with \textit{Gaia} DR2 astrometry to obtain a probable sample of massive stars and analyze their spatial and kinematic distribution in the solar neighborhood. Pantaleoni González et al. identify a prominent overdensity of OB stars \citep[first proposed by][]{Morgan1953} extending 2 kpc in length and bridging the Local Arm and Perseus arm toward the Milky Way’s outer galaxy, which they call the Cepheus spur (shown in purple in Figure \ref{fig:synthesis}). The Cepheus spur contains the prominent OB associations Cep OB4, Cam OB1, Aur OB1, and Gem OB1. Analysis of the vertical distribution of its massive stars indicates that the Cepheus spur has an anomalous average z-height, extending 50 - 100 pc above the Galactic plane. Analysis of its proper motions indicates that the Cepheus spur is kinematically distinct, with a common peculiar motion with respect to neighboring distributions of OB stars. Pantaleoni González et al. suggest that the Cepheus spur is likely related to the Radcliffe Wave, as the spur touches the Wave at one end, coinciding with a region of enhanced star formation at the crest of the wave. 

Complementary maps of the distribution of OB stars derived from \textit{Gaia} astrometry and photometry show many of the same defining features as the Pantaleoni González et al. map \citep[see e.g.][]{Xu2018, Xu2021, Chen2019_OB} used to discover the Cepheus spur. For example, all maps show kiloparsec-scale overdensities associated with the Local and Sagittarius arms --- particularly aligned with these arms’ prominent sub-features like the Radcliffe Wave and the Sagittarius Spur.

\subsection{\textbf{A New Framework for Connecting Kiloparsec to Parsec Scales}}

Together, the discovery of these kiloparsec-scale structures --- containing long-studied clouds previously thought to be in relative isolation --- provides a new framework for testing the influence of galactic dynamics in cloud formation and evolution. 

For example, new simulations featuring resolved cloud populations \citep{Duarte-Cabral2016, Kim2017, Kim2020, Seifried2017, Smith2019, Walch2015, Jeffreson2020} provide the key context for understanding relationships between kiloparsec-scale structures and the clouds and young stars they host, offering the potential to explain how spiral shocks and differential rotation affect the large-scale accumulation of gas in spiral galaxies. Many of these simulations offer the same dynamic range now available in the solar neighborhood. For example, the Latte suite of FIRE-2 simulations of Milky Way-like galaxies can track massive molecular clouds with $\approx 1$ pc spatial resolution and $\approx\rm 1 \; Myr$ temporal resolution, and has been shown to reproduce the observed distribution of molecular cloud masses seen in the Milky Way \citep{Benincasa2018}. Detailed work has also been done to map the ensemble of giant molecular clouds formed consistently in the FIRE-2 simulations to their stellar clusters using new cluster formation models, rendering detailed constraints on the cluster mass, age, and metallicity as a function of both space and time \citep{Grudic2022}. The FIRE cosmological simulations also offer synthetic \textit{Gaia} surveys for select Milky Way-like galaxies, facilitating further comparison between simulations and the observations presented in this review \citep{Sanderson2020}. 

Beyond simulations, JWST is also opening up a new era of resolved cloud populations, offering the same high-resolution ``face-on" view now attainable in the solar neighborhood. In particular, the PHANGS-JWST treasury program \citep{Lee2021, Lee2022} will target nineteen nearby ($\rm d \lesssim 20 \; Mpc$) galaxies with MIRI and NIRCam imaging with a resolution ranging from $\rm 5-50 \; pc$. In combination with UV-optical data on these galaxies' young stellar clusters from Hubble \citep[PHANGS-HST;][]{Lee2022}, spectroscopy of their HII regions from MUSE \citep[PHANGS-MUSE;][]{Santoro2022}, and CO observations of their molecular clouds with ALMA \citep[PHANGS-ALMA;][]{Leroy2021}, the PHANGS-JWST treasury program will enable a complete inventory of star formation activity across the electromagnetic spectrum in these galaxies. 

Given the diversity of tracers now capable of probing the 3D structure of our solar neighborhood, including 3D dust, young stellar objects, OB stars, and open clusters, future efforts to contextualize one tracer in the context of complementary tracers should allow better comparisons to both these extragalactic observations of face-on spiral galaxies and synthetic galaxies from numerical simulations. 

\section{\textbf{CONCLUSIONS}} \label{conclusions}

The deluge of data from the \textit{Gaia} satellite\index{Gaia satellite} has already revolutionized our understanding of the distribution of the gas and young stars in the Milky Way and has revealed a new Galactic solar neighborhood with unmatched detail. The previous model for the distribution of gas and young stars in the local kiloparsec (the Gould’s Belt) is now in crisis, and this structure appears to be made up of parts of previously unrecognized kiloparsec-long gas structures, the Radcliffe Wave and the Split. At the same time, \textit{Gaia} challenges the current description of the structure, pitch angles, location, and mass of the Galactic spiral arms. These unexpected Galactic-scale features can also be resolved down into their individual star-forming complexes in 3D on parsec scales for the first time, opening up a new window for understanding the multi-scale physics of star formation. 

The main conclusions of this chapter are: 

\begin{itemize}

\item We can now begin to reconstruct the 3D spatial distribution of dense gas and young stars at $\approx$ 1 pc spatial resolution using a combination of 3D dust mapping and astrometry of young stars obtained from \textit{Gaia} and VLBI. Detailed studies of individual star-forming regions show that most clouds are highly structured, with filamentary and/or sheet-like morphologies. Many clouds show evidence of being shaped by recent stellar and supernova feedback events, and either contain sub-cavities or are draped on the edge of large-scale superbubbles. 

\item We are starting to be able to measure the 3D motion of cloud complexes using embedded young stars as proxies for the motion of their parental gas. Variations in the 3D structure of star-forming regions map to variations in their dynamics. Several regions show evidence of sequential star formation via gradients in their ages, kinematics, and spatial distribution, suggesting that feedback may shape the 3D dynamics of the local gas and that large-scale triggering may give rise to many of the local young stellar populations.

\item 3D dust mapping of the interstellar medium out to 2 kpc and beyond reveals that molecular clouds are not isolated objects. Instead, clouds are connected and embedded in narrow kpc-long gas structures with widths of the order of $\approx$ 100 pc. These long and narrow structures resemble the ``dust lanes” seen in nearby galaxies.

\item There are two main families of molecular clouds and star-forming regions in the solar vicinity, corresponding to two of these narrow kpc-long gas structures 1) the ``Radcliffe Wave" family of clouds containing Canis Major, Monoceros R2, Orion, Taurus, Perseus, Cepheus, North America, and the Cygnus regions, and 2) the ``Split" family of clouds, including the Sco-Cen related clouds (Ophiuchus, Lupus, Corona Australis, Chameleon), Serpens, and Aquila. These 3D dust features --- and similar features detected in the distribution of OB stars and young stellar clusters --- possess higher pitch angles than predicted by classical spiral structure theory. 

\item Regarding the fundamental question of how clusters and associations dissolve to form the Galactic field, \textit{Gaia} has opened the critical $\rm 10-100 \; Myr$ window on the study of stellar populations, allowing the identification of young stellar populations without infrared excesses near star-forming complexes and young stellar populations far from interstellar gas.

\item Many open clusters, long thought to be roughly spherical structures, appear to be cigar-shaped cluster coronae, with most of the stars residing beyond the cluster core.

\end{itemize}

Ultimately, \textit{Gaia} has enabled for the first time a coherent 3D spatial model of young stars, gas, and dust at a large dynamic range. This 3D spatial distribution can be combined with 3D stellar motions to probe the dynamical state and star formation history of the solar neighborhood in 6D. Yet this field is still in its infancy, with more data and insights still to come. The next \textit{Gaia} data release --- \textit{Gaia} DR4 --- is expected after late 2025 and will be based on 66 months of data collection. \textit{Gaia} DR4 will contain full astrometric, photometric, and radial velocity catalogs, all variable and non-single-star solutions, as well as epoch and transit data for all sources (including all available $\rm BP-RP$ and RVS spectra) \citep{Collaboration2022}. Parallaxes will improve by a factor of $1.4\times$ and proper motions by a factor of $2.8\times$ compared to \textit{Gaia} DR3 \citep{gaiastats}. The final \textit{Gaia} data release, \textit{Gaia} DR5, should include $> 10$ years of observations and release after 2030 \citep{Collaboration2022}. Compared to \textit{Gaia} DR4, parallaxes, photometry, and radial velocities should improve by $1.4\times$ and proper motions by a factor of $2.8\times$ in \textit{Gaia} DR5 \citep{gaiastats}. Accurate tangential velocities should also be available for a $>22\times$ larger volume. This combination of future \textit{Gaia} data releases with the wealth of photometric and spectroscopic surveys on the horizon --- including SDSS-V and LSST \citep{SDSSV,LSST} --- will only continue to enhance our models of the 6D distribution of gas and young stars, paving the way for a unified understanding of how star formation is mediated by physical processes occurring across vastly different scales in both time and space. 

\bigskip

\noindent\textbf{Acknowledgments} We thank Eric Mamajek and a second anonymous referee for excellent comments that substantially improved this chapter. Catherine Zucker acknowledges that support for this work was provided by NASA through the NASA Hubble Fellowship grant \#HST-HF2-51498.001 awarded by the Space Telescope Science Institute (STScI), which is operated by the Association of Universities for Research in Astronomy, Inc., for NASA, under contract NAS5-26555. Phillip Galli acknowledges financial support from São Paulo Research Foundation (FAPESP) under grants 2020/12518-8 and 2021/11778-9.

\bigskip

\bigskip
\parskip=0pt

\bibliographystyle{pp7}
\bibliography{Ch2.bib}

\end{document}